\documentclass{aa}  
\usepackage{graphicx}
\usepackage{txfonts}
\usepackage{natbib}
\usepackage{longtable,lscape}

\newcommand{\Hm}{H$_2$}
\newcommand{\Bg}{Br$\gamma$}

\newcommand{\HII}{\ion{H}{ii}}

\newcommand{\um}{$\mu$m}
\newcommand{\iras}{IRAS\,19410$+$2336}
\newcommand{\spitzer}{{\it Spitzer}}

\begin{document}
   \title{Spatially resolved near-infrared spectroscopy of the massive star-forming region IRAS\,19410+2336}

   \author{N.~L. Mart\'{\i}n-Hern\'{a}ndez
          \inst{1}
          \and
	  A. Bik
	  \inst{2}
	  \and 
	  E. Puga
	  \inst{3,4}
	  \and
	  D.~E.~A. N\"{u}rnberger
	  \inst{5}
	  \and
	  L. Bronfman
	  \inst{6}
          }

   \offprints{N.L.\,Mart\'{\i}n-Hern\'{a}ndez, leticia@iac.es}

   \institute{Instituto de Astronom\'ia de Canarias, V\'ia L\'actea s/n,
             38205 La Laguna, Spain
             \and
	     Max Planck Institute for Astronomy, K\"{o}nigstuhl 17, 69117 Heidelberg, Germany
	     \and
	     Instituut voor Sterrenkunde, Katholieke Universiteit Leuven, Celestijnenlaan 200D, 
	     3001 Leuven, Belgium
	     \and
	     Departamento de Astrof\'isica Molecular e Infrarroja, Instituto de Estructura de la
	     Materia, CSIC, Serrano 121, 29006 Madrid, Spain 
	     \and 
	     European Southern Observatory, Casilla 19001, Santiago 19, Chile
	     \and
	     Departamento de Astronom\'ia, Universidad de Chile, Casilla 36-D, Santiago, Chile
	     }

   \date{Received; accepted}

 
  \abstract
   {}  
   {\iras\ is a young massive star forming region with an intense outflow activity. Outflows are frequently studied in the near-infrared (NIR) since the \Hm\ emission in this wavelength range often traces the shocked molecular gas. However, the mechanisms behind the \Hm\ emission detected in \iras\ have not been clarified yet. We present here spatially resolved NIR spectroscopy which allows us to verify whether the \Hm\ emission originates from thermal emission in shock fronts or from fluorescence excitation by non-ionizing UV photons. Moreover, NIR spectroscopy also offers the possibility of studying the characteristics of the putative driving source(s) of the \Hm\ emission by the detection of photospheric and circumstellar spectral features, and of the environmental conditions (e.g. extinction).}
   {We obtained long-slit, intermediate-resolution, NIR spectra of \iras\ using LIRIS, the NIR imager/spectrographer mounted on the 4.2\,m William Herschel Telescope. As a complement, we also obtained $J$, $H$ and $K_s$ images with the Las Campanas 2.5\,m Du Pont Telescope, and archival mid-infrared (MIR) \spitzer\ images at 3.6, 4.5, 5.8 and 8.0\,\um.}
   {We confirm the shocked nature of the \Hm\ emission, with an excitation temperature of about 2000~K, based on the analysis of relevant \Hm\ line ratios, ortho-to-para ratios and excitation diagrams. We have also identified objects with very different properties and evolutionary stages in \iras. 
   The most massive source at millimeter wavelengths, mm1, with a mass of a few tens of solar masses, has a bright NIR (and MIR) counterpart.  This suggests that emission --probably coming through a cavity created by one of the outflows present in the region, or from the outflow cavity itself-- is leaking at these wavelengths. 
   The second most massive millimeter source, mm2, is only detected at $\lambda \gtrsim 6$\,\um, suggesting that it could be a high-mass protostar still in its main accretion phase. The NIR spectra of some neighboring sources show CO first-overtone bandhead emission which is associated with neutral material located in the inner regions of the circumstellar environment of YSOs. }
   {}


  \titlerunning{Spatially resolved NIR spectroscopy of IRAS\,19410+2336}
  
  \maketitle
   
%

\section{Introduction}

\iras\ is a young massive star forming region which has been extensively studied in a series of papers during the last few years. It is considered as a prime example of the complexity of high-mass star formation due to the intense outflow activity ongoing there. 
The 1.2 mm dust continuum map of \iras\ obtained with the IRAM 30\,m telescope at low spatial resolution reveals two massive adjacent star forming clumps roughly aligned in a north-south direction and separated by $\sim$53\arcsec\ \citep[][]{beuther02c}. The southern clump, with a mass of about 840~$M_\odot$, is more massive than the northern clump, with a mass of 190~$M_\odot$. Observations in CO(2--1) with 11\arcsec\ angular resolution show that each clump is associated with a bipolar molecular outflow in the east-west direction \citep{beuther02d}. The southern clump is resolved into at least four cores in the 2.6\,mm continuum using the Plateau de Bure Interferometer, whereas the northern clump is resolved into two cores \citep{beuther03}. 
These authors disentangle at least seven (possibly even nine) bipolar outflows in this region with accretion rates of the order of $10^{-4} M_\odot$\,yr$^{-1}$, sufficiently high to overcome the radiation pressure and form massive stars via continuous disk-accretion processes. 
Higher spatial resolution observations at 1.3\,mm, also with the Plateau de Bure Interferometer, show that these two clumps split up into a even larger number of sub-structures, with at least 12 cores in each one \citep{beuther04}. These cores have masses ranging between 1.7 and 25\,M$_\odot$, and visual extinctions of up to 1000 magnitudes.
 
At the very center of the southern clump and coinciding with the most prominent of the mm sources, \cite{beuther02b} detected H$_2$O and \ion{Class}{ii} CH$_3$OH masers, which are considered to be closely associated with the earliest stages of massive star formation. Only at this position is there evidence for a weak 1~mJy unresoved (at angular resolution of 0\farcs7) cm continuum source \citep{sridharan02}.

\cite{beuther02} reported the detection of hard X-ray emission from a number of point sources in the vicinity of \iras. They conclude that this X-ray emission is due to intermediate-mass pre-main-sequence Herbig Ae/Be stars or their precursors. 

So far, only the kinematic distance to this object is known and there exists a distance ambiguity, with a near distance of 2.1~kpc and a far distance of 6.4~kpc \citep{sridharan02}. However, authors like \cite{beuther02} consider this source to be located at its near kinematic distance because the derived outflow parameters are unreasonably high if the far kinematic distance is assumed \citep[see also][]{beuther02d}. Other arguments that strengthen the selection of the near kinematic distance are the following. (1) The CS(2--1) profile for \iras\ \citep{bronfman96} is fairly narrow (with a FWHM of 3.1~km/s) as compared with other sources that also have a complex outflow structure, for instance IRAS\,18507$+$0121, with a linewidth of 6.8~km/s at a distance of 3.5~kpc \citep{shepherd07}. At a distance of 2.1~kpc, the 40\arcsec\ beam of these observations is probably detecting only the mm clump and not the lobes observed in CO(2--1). At the far distance of 6.4~kpc, however, the red and blue lobes of the outflow would have been detected as wings in the CS(2--1) profile. (2) \cite{bronfman96} observed four IRAS point sources with far-infrared colors typical of ultracompact \HII\ regions, including \iras, which have very similar (LSR) velocities. These sources are within a galactic extent of about 1.7\degr, or about 60~pc at a distance of 2.1~kpc. This is the typical size of a giant molecular cloud (GMC), and it is likely that these four sources belong to the same GMC. At the far distance of 6.4~kpc, on the contrary, there would have to be a series of GMCs with the same velocity, or a single GMC with an extent of 180~pc, which is unlikely. (3)  \cite{watson03} observed \iras\ in the H110$\alpha$ and H$_2$CO lines in order to solve the distance ambiguity. No emission was detected in H110$\alpha$, but they detected H$_2$CO in absorption with a (LSR) velocity of 22.6~km/s and a linewidth of 4.7~km/s. No absorption is detected at larger velocities. 

We thus assume a distance of 2.1~kpc. At this distance, the infrared luminosity of \iras\ is $10^4$~L$_\odot$.

Outflows are frequently studied in the near-infrared (NIR), since the \Hm\ emission in this range often traces the hot \mbox{($<2000$~K)} shocked molecular gas. In both low-luminosity and high-luminosity outflows, the \Hm\ emission delineates single or multiple bow-shaped features associated with the leading edge of the CO emission and, in some cases, also traces collimated jets and cavity walls \cite[e.g. see the review by][]{richer00}.
\cite{beuther03} mapped the \Hm\ emission around \iras\ with the 3.5\,m telescope on Calar Alto and used this information (combined with high-spatial resolution CO observations) in their interpretation of the outflows occurring in this region. However, the excitation mechanism behind this \Hm\ emission has not yet been determined. 
With this aim in mind, we obtained long-slit, intermediate-resolution, NIR spectra of \iras\ using LIRIS, the NIR imager/spectrographer mounted on the 4.2\,m William Herschel Telescope (WHT). These observations help us verify whether the \Hm\ emission is produced by thermal emission in shock
fronts or by fluorescence excitation by non-ionizing UV photons. These mechanisms can be distinguished since they preferentially populate different levels producing different \Hm\ line ratios. 
NIR spectroscopy also offers the possibility of studying the characteristics of the putative driving source(s) of the \Hm\ emission by the detection of photospheric and circumstellar spectral features. Moreover, it allows us to investigate environmental conditions (e.g. extinction) towards \iras. 

As a complement, we also obtained $J$, $H$ and $K_s$ images with the Las Campanas 2.5\,m Du Pont Telescope. Combined with archival \spitzer\ data at 3.6, 4.5, 5.8 and 8.0~\um, obtained by the GLIMPSE Legacy Science Program \citep{benjamin03}, these images are helpful to detect optically obscured, deeply embedded young stars and protostellar candidates, excess emission from their circumstellar matter and, at earlier stages, from their infalling envelopes.

This paper is structured as follows. Section~\ref{sect:data} describes the observations and data reduction, and Sect.~\ref{sect:results} presents the photometric and spectroscopic results. Finally, Sect.~\ref{sect:conclusions} presents the discussion and main conclusions.

\section{Observations and data reduction}
\label{sect:data}

\subsection{Imaging}

$J$, $H$ and $K_s$ images were taken with the Las Campanas 2.5\,m Du Pont Telescope. This telescope is equiped with a NIR camera  \citep{persson92} using a Nicmos-4 $256\times256$ pixel HgCdTe array detector.  The observations of \iras\ were taken on 1996 May 26. Integration times of 5 seconds per frame were used. A dither pattern using 20--25\arcsec\ offsets were applied resulting in an on-source integration time of 35 seconds.

\begin{figure*}
    \begin{minipage}[c]{0.5\textwidth}
      \centering \resizebox{\hsize}{!}{\includegraphics{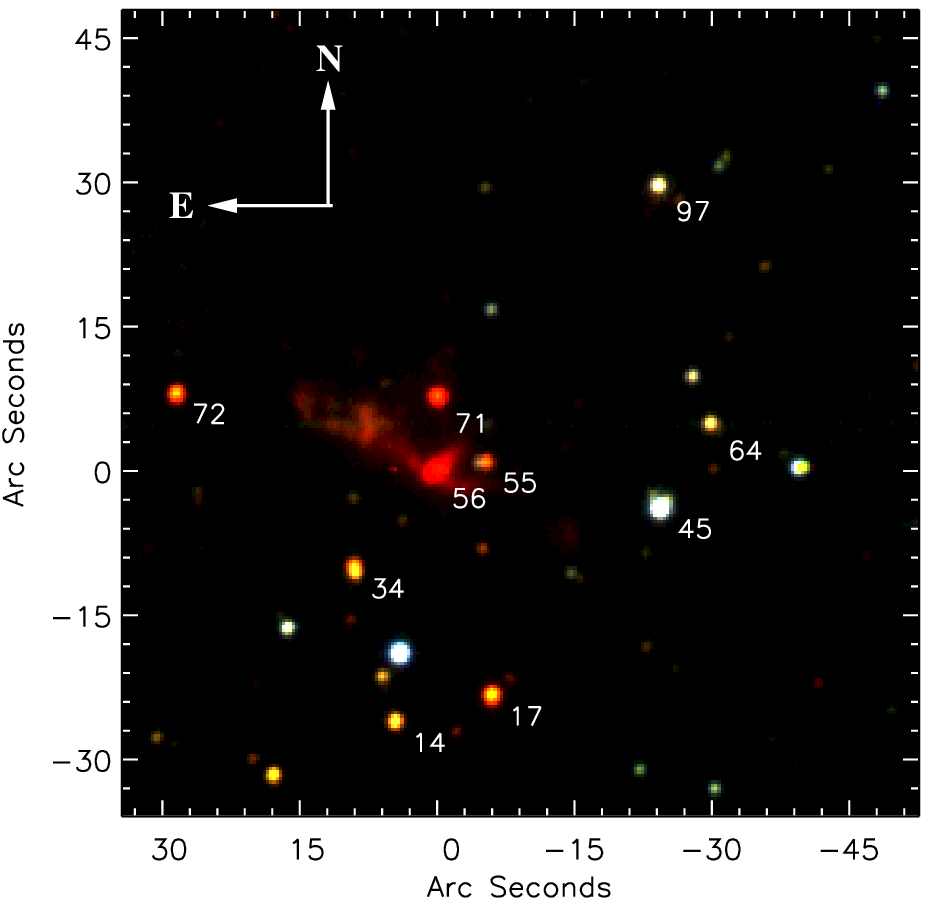}}
    \end{minipage}%
    \begin{minipage}[c]{0.5\textwidth}
      \centering \resizebox{\hsize}{!}{\includegraphics{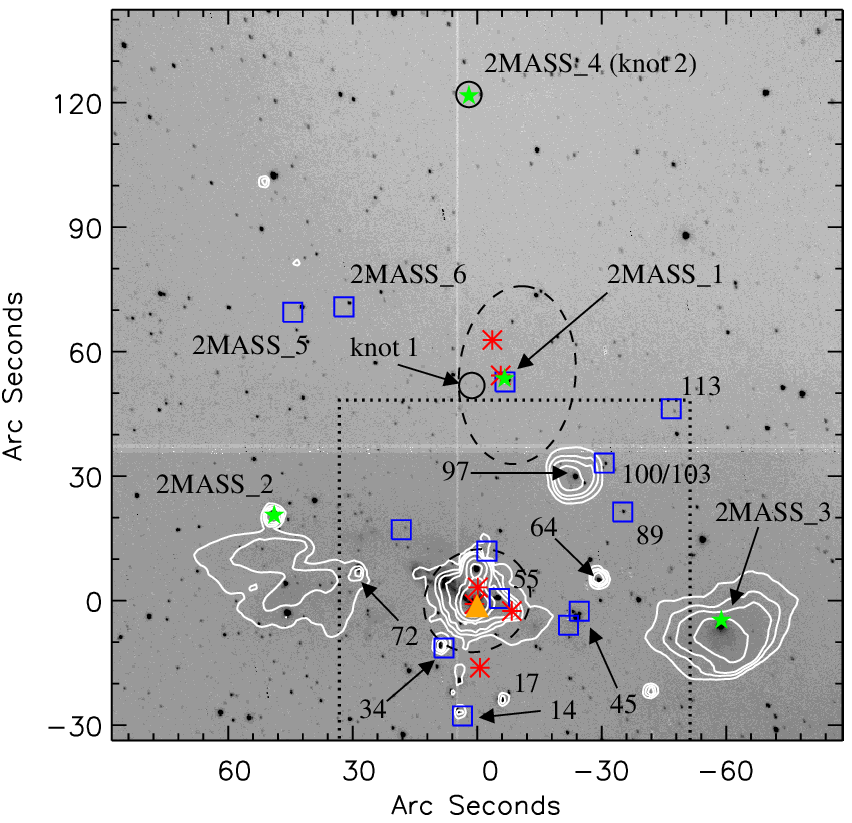}}
    \end{minipage}
       \caption{
       {\it (Left)} Las Campanas color composite image of \iras\ where $J$ is blue, $H$ 
       is green and $K_s$ is red. Stars which are discussed in the text are labeled.
       {\it (Right)} In grey-scale we show the broad-band ($K_s$ filter)
       acquisition image of \iras\ obtained with LIRIS. A dotted box outlines the field-of-view covered by the Las Campanas $JHK_s$ data.
       The black small circles indicate the positions of the H$_2$ knots 1 and 2 detected in one of the two LIRIS slits (slit \#1).
       The large dashed circles mark the positions and approximate sizes of the two massive star-forming clumps detected at 1.2\,mm with the IRAM 30\,m telescope.
       The white contours outline the {\it Spitzer} 5.8 \um\ 
       emission (more detailed MIR images of the central source are shown in Fig.~\ref{fig:irac}).
       The red asterisks mark the positions of the mm sources identified by \cite{beuther03}.
       The orange filled triangle marks the position of the 
       cm continuum source associated with H$_2$O and CH$_3$OH maser emission
       \citep{beuther02b}. 
       Blue squares mark the X-ray sources observed by \cite{beuther02}.      
       The green stars are 2MASS point sources of interest.
       The (0\arcsec,~0\arcsec) position in both images corresponds to 
       R.A. (J2000.0)=19$^{\rm h}$43$^{\rm m}$11\fs18 and 
       Dec (J2000.0)=23\degr44\arcmin04\farcs2 [see the electronic edition of the Journal for a color version of this figure].}
        \label{fig:im}
  \end{figure*}

A color composite image of \iras\ is presented in the left panel of Fig.~\ref{fig:im}. Relevant stars are indicated.

The measured  FWHM is 1$\farcs$2, 1$\farcs$1 and 1$\arcsec$ in the $J$, $H$ and $K_s$ band, respectively.
We obtained the photometry of IRAS19410+2336 using the standard IRAF\footnote{IRAF is distributed by the National Optical Astronomy Observatory, which is operated by the Association of Universities for the Research in Astronomy, Inc., under cooperative agreement with the National Science Foundation (http://iraf.noao.edu/)} DAOPHOT PSF-fitting routines on the final mosaicked frames.  We constructed  a semi-analytical PSF  from 10 isolated stars in the field for each  band. The PSF postage stamp radius was set to 3$\farcs$15  while the source fitting radius was 1$\arcsec$.
2MASS zeropoints were estimated comparing the fluxes  of 14 bright sources --selected from  other fields observed during this same night-- with their brightnesses in the 2MASS All-Sky Catalogue of Point Sources. We did not apply any color correction. A total of 47, 67 and 110 sources were extracted in the J, H and K$_s$  band, respectively.

The electronic Table~1 lists the final $J$, $H$ and $K_s$ magnitudes in the 2MASS photometric system for the stars in the field-of-view depicted in the left panel of Fig.~\ref{fig:im}. The final uncertainties are the result of propagating the errors coming from the instrumental magnitudes and those from the linear fits used in the magnitude transformation. Limiting magnitudes are around 17.5 mag in $J$, 17 mag in $H$ and 16.5 mag in $K_s$.

Astrometry was performed by matching the positions of stars in common in our images and in the 2MASS survey catalogue. The astrometry accuracy is of the order of 0\farcs7.

\begin{figure}
    \begin{minipage}[c]{0.5\textwidth}
      \centering \resizebox{\hsize}{!}{\includegraphics{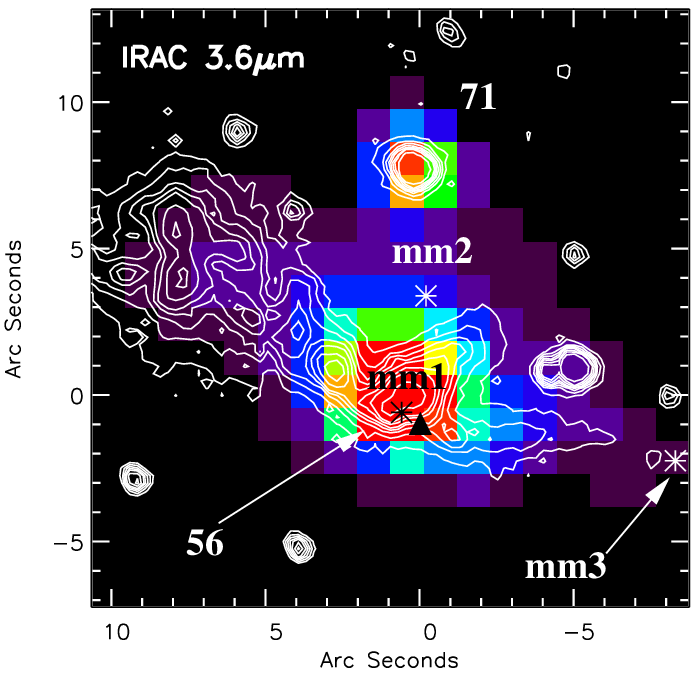}}
    \end{minipage}\\
    \begin{minipage}[c]{0.5\textwidth}
      \centering \resizebox{\hsize}{!}{\includegraphics{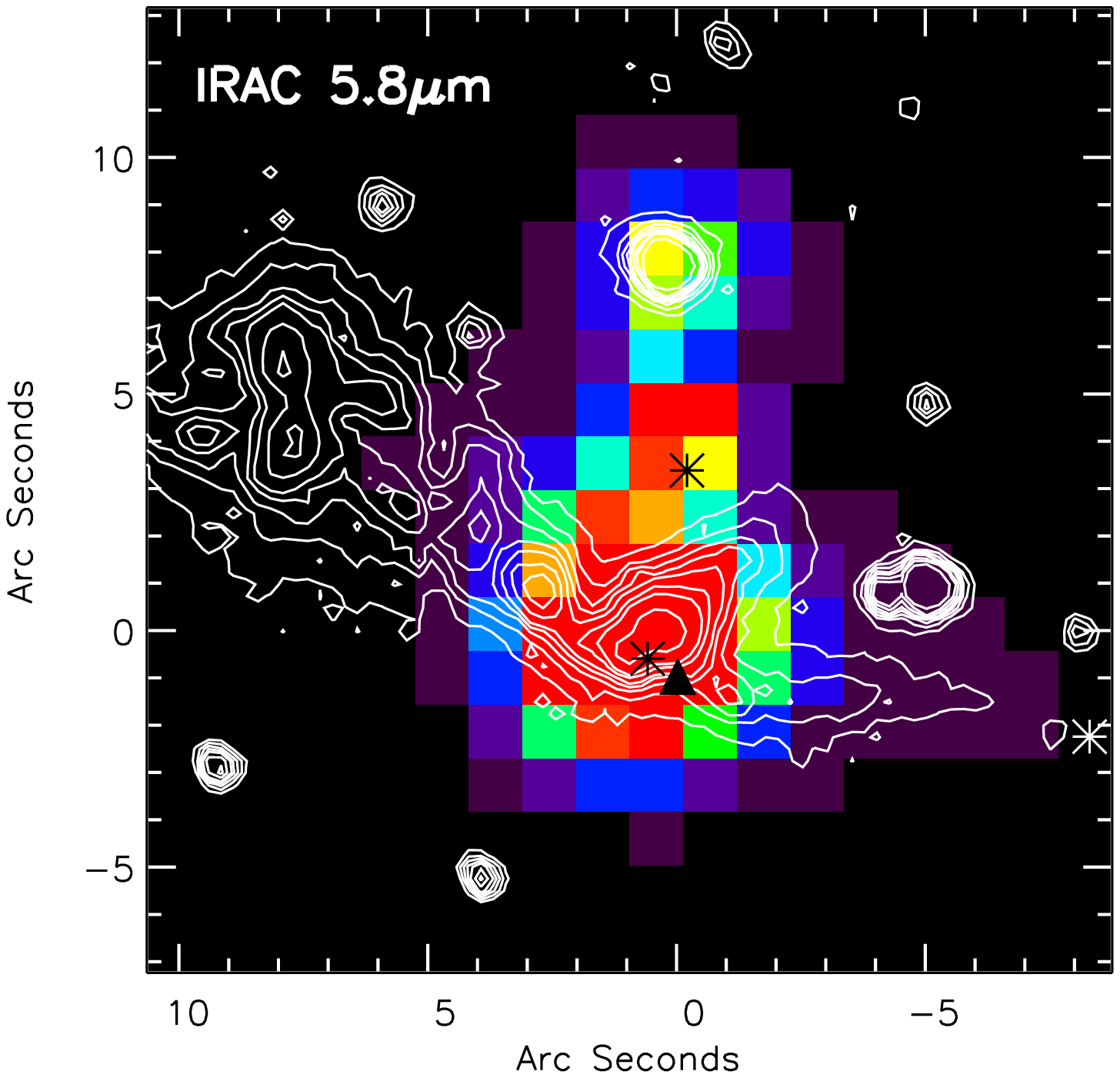}}
    \end{minipage}
       \caption{\spitzer/IRAC 3.6 and 5.8~\um\ images of the central part of \iras. White contours represent the $K_s$ emission. The mm sources identified by \cite{beuther03} are indicated by asterisks. The triangle marks the position of the cm continuum source. In these figures, north is up and east is to the left.
       The (0\arcsec, 0\arcsec) position corresponds to 
       R.A. (J2000.0)=19$^{\rm h}$43$^{\rm m}$11\fs18 and 
       Dec (J2000.0)=23\degr44\arcmin04\farcs2  [see the electronic edition of the Journal for a color version of this figure].}
        \label{fig:irac}
  \end{figure}

Archival \spitzer\ images at 3.6, 4.5, 5.8 and 8.0~\um\ are also used. The contours of the 5.8~\um\ data are overplotted in the right panel of Fig.~\ref{fig:im}. Fig.~\ref{fig:irac} shows a more detailed comparison between the \spitzer\ and the $K_s$ imaging data. Saturation affected the brightest components in the 4.5 and 8.0~\um\ band images, and so they are not shown. Saturation limits at 4.5 and 8.0~\um\ are, respectively, 200 and 740 mJy for a frame time of 2 seconds.
\spitzer/IRAC magnitudes in the 3.6, 4.5, 5.8 and 8.0\,\um\ filters are also listed in Table~1. They have been obtained from the GLIMPSE Archive \citep[e.g.][]{benjamin03}, available through the NASA/IPAC Infrared Archive\footnote{http://irsa.ipac.caltech.edu}. 3--$\sigma$ detection limits are 15.5, 15.0, 13.0 and 13.0 mag in the  3.6, 4.5, 5.8 and 8.0~\um\ bands, respectively.

  \begin{figure}
     \centering \resizebox{\hsize}{!}{\includegraphics{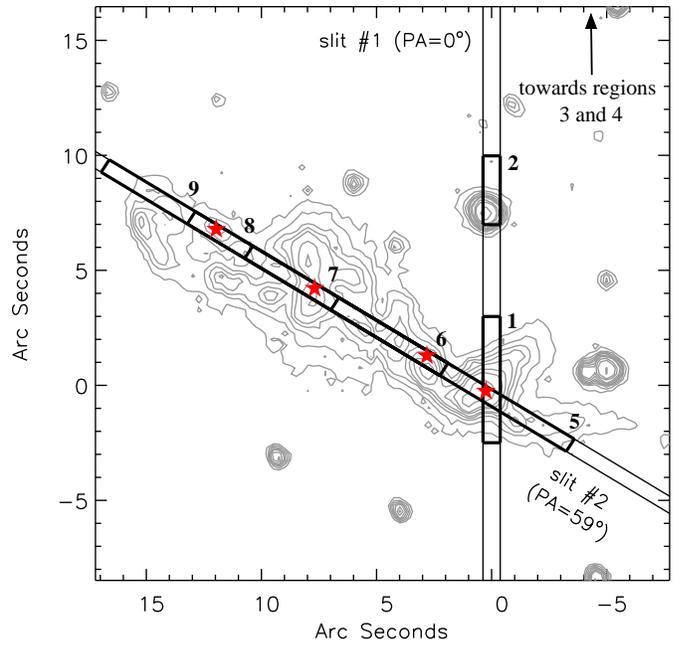}}
     \caption{Contours of the broad-band ($K_s$ filter) acquisition 
       image of \iras\ obtained with LIRIS.
       The positions of the slits and the distinct regions analysed in 
       the text 
       are marked. Regions 3 and 4, which correspond respectively to H$_2$ knots 1 and 2 (see the right panel of Fig.~\ref{fig:im}), are 50\arcsec\ and 119\arcsec\ north of nr56. The red stars indicate, from right to left, the positions of sources nr56, nr56b, nr56c and nr56d. In this figure, north is up and east is to the left.
       The (0\arcsec, 0\arcsec) position corresponds to 
       R.A. (J2000.0)=19$^{\rm h}$43$^{\rm m}$11\fs18 and 
       Dec (J2000.0)=23\degr44\arcmin04\farcs2 [see the electronic edition of the Journal for a color version of this figure].}
     \label{fig:slits}
  \end{figure}

\subsection{Spectroscopy}

NIR spectra in the 1.9--2.4~\um\ wavelength range (using the {\it mrk} grism) were obtained during the nights of 2006 August 31 and September 1 and 2 at the 4.2\,m WHT using LIRIS, a long-slit intermediate-resolution infrared spectrograph \citep{acosta03,manchado04}.
LIRIS is equipped with a Rockwell Hawaii $1024\times1024$ HgCdTe array detector. The spatial scale is 0\farcs25 pixel$^{-1}$, and the slit width used during the observations was 0\farcs75, allowing a spectral resolution $R=\lambda/\Delta\lambda\sim 1000$. The slit (of length 4\farcm2) was oriented along P.A.$=0\degr$ (slit \#1) and 59\degr\ (slit \#2) covering the brightest nebular spots (see Fig.~\ref{fig:slits}). Weather conditions were relatively good, although with sparse cirrus. The infrared seeing during our observations varied between 0\farcs7 and 1\farcs8 measured from the FWHM of the standard star spectra.

Observations were performed following an ABBA telescope-nodding pattern, placing the source in two positions along the slit. Offsets were 15\arcsec\ for slit \#1 and 25\arcsec\ for slit \#2. Individual exposures were taken with integration times of 300 seconds, adding up to a total on-source integration time of 80 minutes for slit \#1, and 60 minutes for slit \#2. In order to obtain the telluric correction and the flux calibration, the nearby AO\,V stars HIP\,95487 and HIP\,95560 were observed with the same configuration. The data were reduced following standard procedures for NIR spectroscopy, using IRAF and LIRISDR, the LIRIS data reduction package. After the flat-field correction, consecutive AB pairs of two-dimensional spectra were subtracted to remove the sky background. The resulting frames were then co-added to provide the final spectrum. Sky lines present in the science raw data were used to determine the wavelength calibration. The final uncertainty in the wavelength calibration is of the order of 1\AA. The resulting wavelength-calibrated spectra were divided by a normalized reference spectrum to remove the telluric contamination. This normalized reference spectrum was generated from the observed spectra of the A0\,V standard stars, divided by a Vega model spectrum convolved with the actual spectral resolution ($\sim 23$\AA). For this, we used the routine {\it xtellcor\_general} within the Spextool package \citep{vacca03}.
This routine was also used for the flux calibration. We estimate an uncertainty in the flux calibration around 20\% based on the comparison of the various spectra obtained for the standard stars. This large uncertainty probably reflect the fact that sparse cirrus were present during the time of observation.

The contours of the (acquisition) broad-band image of \iras\ obtained in the $K_s$ filter (cut-off wavelengths are 1.990 and 2.310~\um) are shown in Fig.~\ref{fig:slits}. The slits are overplotted. We identify several regions, 1 to 4 in slit \#1 (with P.A.$=0\degr$), and 5 to 9 in slit \#2 (with P.A.$=59\degr$). These regions represent well differentiated parts of the \Hm\ emission. Regions 3 and 4 of slit \#1 correspond to two \Hm\ knots (1 and 2) found at about 50\arcsec\ and 119\arcsec\ north of object nr56 (see the right panel of Fig.~\ref{fig:im}, where their positions are indicated).

Peak positions and fluxes of the lines were measured by fitting a Gaussian profile. These are listed in Table~\ref{table:h2}. The large discrepancies between regions 1 and 5 (both include the source nr56) are due to the different size and orientation of the extraction apertures.
A line is defined as being detected if its peak intensity exceeds the rms noise level of the local continuum (typically around $4.5\times10^{-16}$~W\,m$^{-2}$\,\um$^{-1}$) by at least a factor of~3.

\section{Results}  
\label{sect:results}
  
\subsection{The environment of \iras}  

As mentioned in the introduction, \iras\ is formed by two massive adjacent star-forming clumps \citep[][]{beuther02c}. The southern clump, with a mass of 840\,$M_\odot$, is about four times more massive than the northern clump. The positions and approximate sizes of these unresolved clumps are marked in the right panel of Fig.~\ref{fig:im}. 
The positions of the millimeter sources observed by \cite{beuther03} are also indicated in Fig.~\ref{fig:im}, together with the locations of the unresolved cm continuum emission detected within the southern clump \citep{sridharan02}, and the X-ray sources observed by \cite{beuther02}. 

The elongated emission feature seen in the $K_s$ image coincides with an \Hm\ linear structure  observed by \cite{beuther03}. This elongated \Hm\ feature is strongly associated with the blue-shifted molecular emission of a CO outflow --denominated C by \cite{beuther03}-- and outlines the same morphological structure, indicating that this is a highly collimated outflow. This outflow is probably driven by the millimeter source mm1. This is the brightest mm source, with a mass of a few tens of solar masses \citep{beuther04}, and is associated with the NIR object nr56. It also has a MIR counterpart (see Fig.~\ref{fig:irac}).

  \begin{figure*}[!ht]
    \begin{minipage}[c]{0.5\textwidth}
      \centering \resizebox{\hsize}{!}{\includegraphics{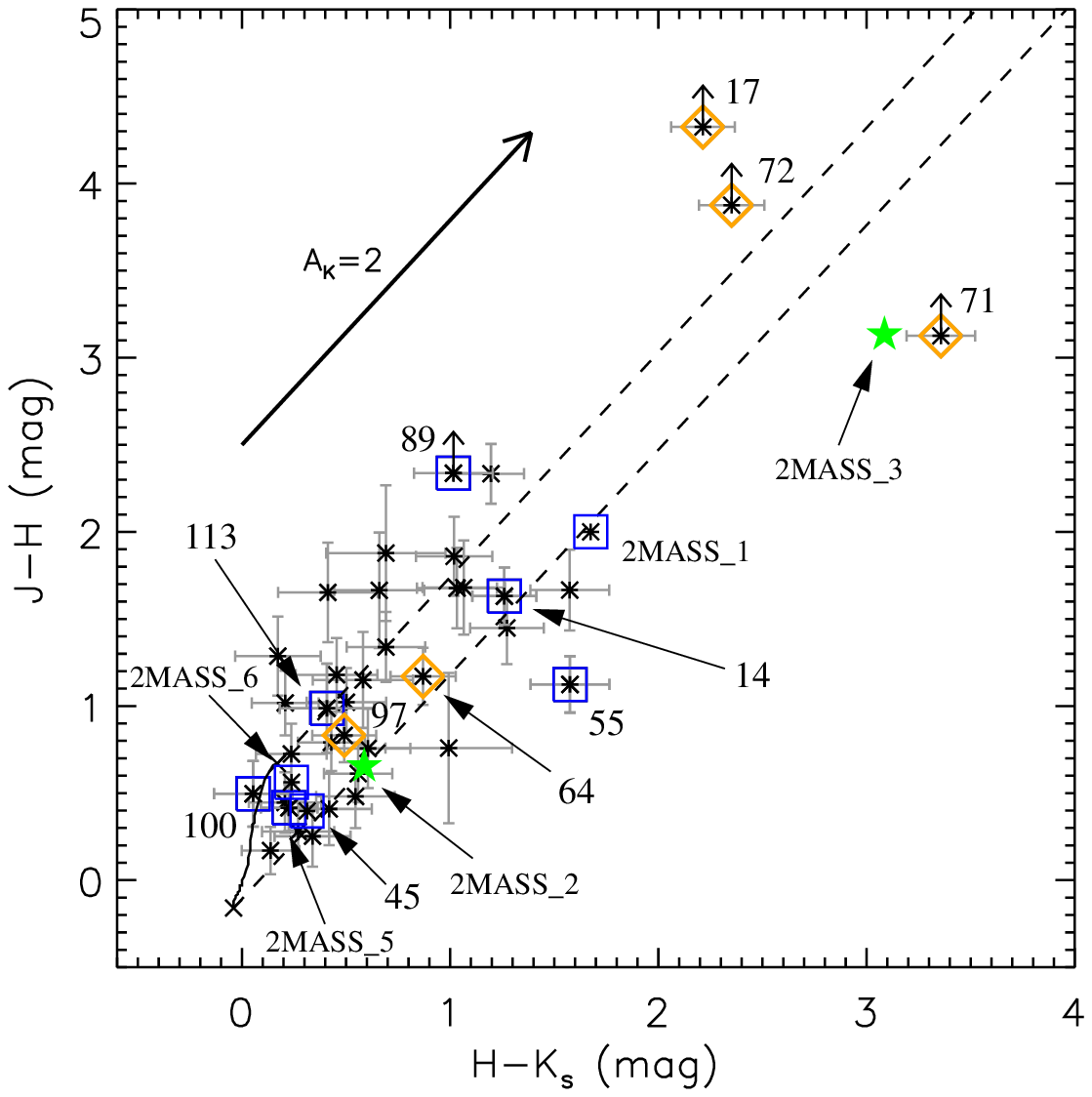}}
    \end{minipage}%
    \begin{minipage}[c]{0.5\textwidth}
      \centering \resizebox{\hsize}{!}{\includegraphics{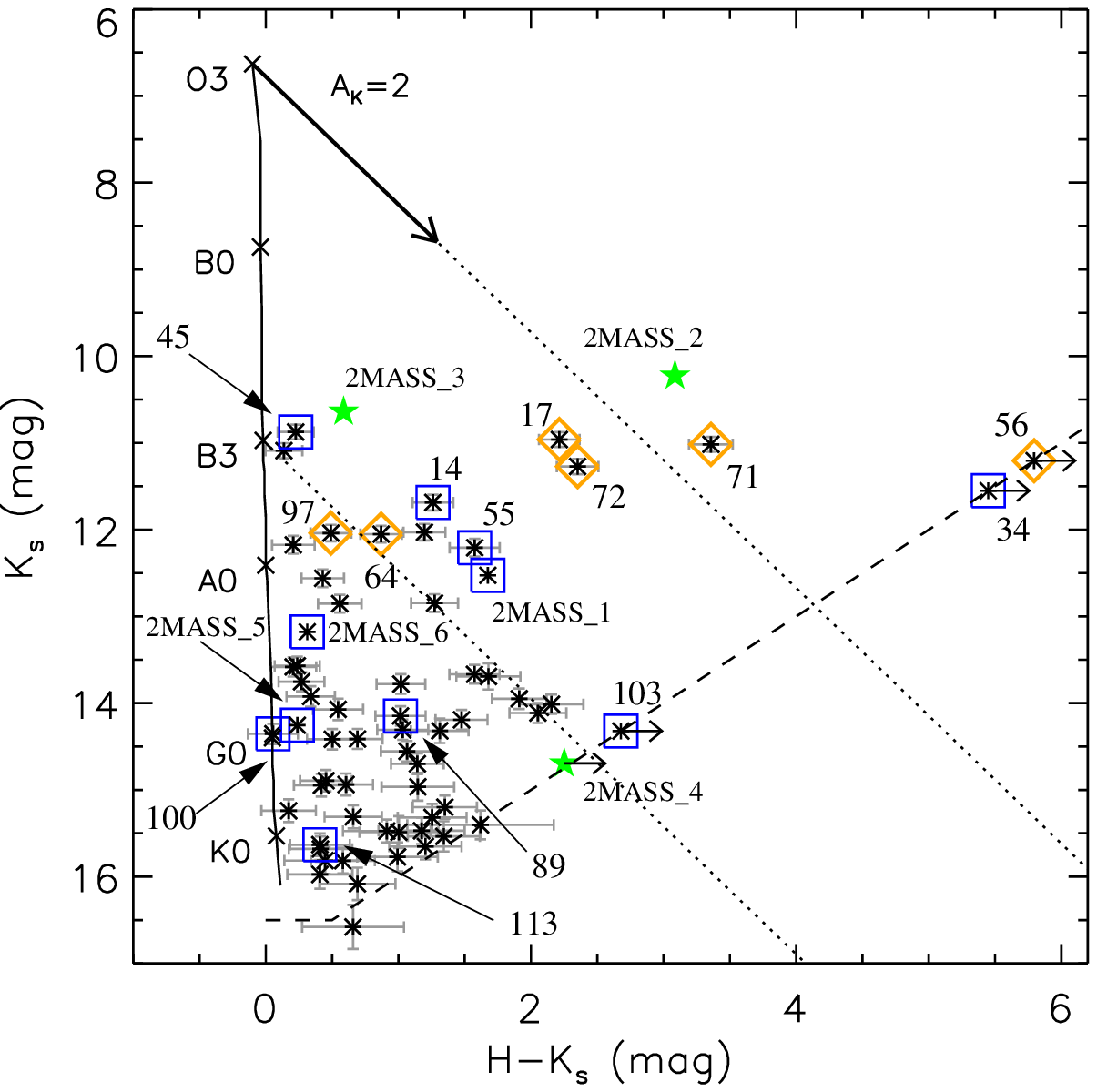}}
    \end{minipage}
       \caption{
       {\it (Left)} NIR color-color diagram of \iras. The solid line represents
       the main sequence from \cite{tokunaga00} and \cite{martins06}. The dashed 
       lines show the reddening band for main sequence colors.
       A reddening
       vector of 2 mag in the $K$-band is represented by a thick arrow. 
       Blue squares indicate sources associated with X-ray emission. Orange diamonds are objects which are  associated with strong 
       MIR emission. Finally, various 2MASS point sources of interest are plotted by green stars. We include 2MASS\_4, the 2MASS point source associated with the H$_2$ knot 2. 
       {\it (Right)} NIR color-magnitude diagram at a distance of 2.1~kpc. Here, the dashed
       lines show our detection limits, and the dotted lines comprise the area where reddened O and early B stars at a distance of 2.1~kpc might be found [see the electronic edition of the Journal for a color version of this figure].}
        \label{fig:phot}
  \end{figure*}

  \begin{figure}[!hb]
      \centering \resizebox{\hsize}{!}{\includegraphics{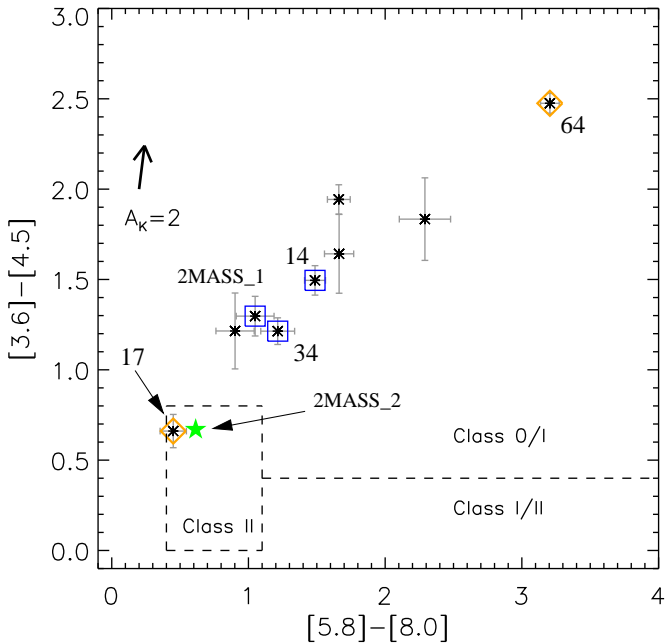}}
       \caption{\spitzer/IRAC color-color diagram for sources with identifications in all four IRAC bands. The thick vector on the left shows $A_K=2$. We use the same symbols here as in Fig.~\ref{fig:phot}. The dashed lines and labels for different types of protostars are based on the classification of \cite{megeath04} [see the electronic edition of the Journal for a color version of this figure].}
    \label{fig:cc:spitzer}
  \end{figure}

Another object of interest is source nr71, which also has a MIR counterpart detected by \spitzer, although weaker, and has no counterpart at millimeter wavelengths. Finally, the millimeter source mm2, which is nearly as massive as mm1 \citep{beuther04}, is associated with MIR emission but remains undetected at  wavelengths shortward of 6~\um.

\subsection{Color-color and color-magnitude diagrams}  
     
The electronic Table~1 contains $JHK_s$ and IRAC magnitudes for 116 stars detected in the $\sim75\arcsec\times75\arcsec$ field around \iras. The two bright red objects seen in the left panel of Fig.~\ref{fig:im} are stars nr56 and nr71, with $K_s$ magnitudes of 11.21 and 11.02, respectively. Source nr56 is only detected in $K_s$.

The color-color ($J-H$ vs $H-K_s$) and color-magnitude ($K_s$ vs $H-K_s$) diagrams are shown in Fig.~\ref{fig:phot}. Error bars are 1--$\sigma$. In each diagram we have plotted the theoretical main sequence at the assumed distance for the region, i.e. 2.1~kpc. Intrinsic colors were taken from \cite{tokunaga00} and \cite{martins06}. Absolute visual magnitudes, $M_v$, were taken from \cite{cramer97}, \cite{houk97} and \cite{martins06}. We also include four sources (2MASS\_1 -- 2MASS\_4) outside the $\sim75\arcsec\times75\arcsec$ field around \iras\ which have available 2MASS photometry. These sources are plotted by green star symbols and identified in the right panel of Fig.~\ref{fig:im}. The large scatter in the color-color diagram could be due to, among other possibilities, the different spatial resolution between 2MASS and the Las Campanas observations, and the presence of blobs of emission with no detected star.

The dotted lines in the color-magnitude diagram comprise the area where reddened O and early B stars at a distance of 2.1~kpc might be found. Such stars would ionize their own \HII\ regions. Among the sources detected in $H$ and $K_s$, several display high extinction appearing as red or orange in the left panel of Fig.~\ref{fig:im}.  These stars are nr14, nr17, nr55, nr64, nr72 and nr97, with $A_K$ in the range  1--4 mag. Stars nr14 and nr55 are associated with two X-ray point sources detected by \cite{beuther02}. Among this group of candidate O and early B type stars are 2MASS\_1, associated with a mm source (mm5), and 2MASS\_3, associated with extended MIR emission. From all these sources, only nr55 and 2MASS\_3 exhibit a NIR infrared excess as can be seen in the color-color diagram. 

Other bright $K_s$ sources are stars nr34, nr56, nr71 and 2MASS\_2. From these, only star nr71 shows a clear NIR excess; stars nr34 and nr56 are not detected in $J$ and $H$. 2MASS\_2, on the other hand, is associated with extended MIR emission.

There are not many stars with a NIR excess even though the X-ray data indicates the presence of Herbig Ae/Be stars in the field around \iras. However, this type of stars might only start showing an infrared excess at wavelengths longward of about 3\,\um.

Fig.~\ref{fig:cc:spitzer} shows a [3.6]-[4.5] vs [5.6]-[8.0] color-color diagram for the sources with identifications in all four IRAC bands, where we use the classification of \cite{megeath04}. The box enclosed by dashed lines shows the domain of Class II objects, i.e. stars with disks. The domain of embedded young stellar objects (YSOs), i.e. of Class 0 and I objects, is also indicated. This diagram is useful to identify infrared excess emission of the observed targets. Sources nr17 and 2MASS\_2 fall within the ``disk domain'', while others, such as 2MASS\_1, nr14, nr34 and nr64, have a larger infrared excess and fall within the domain of embedded YSOs. In particular,  2MASS\_1, nr14 and nr34 have X-ray counterparts. Finally, nr56, with [3.6]-[4.5]$\sim$1.85, could certainly be an embedded YSO. By comparison, nr71, with [5.8]-[8.0]$\sim$1.16, does not have a large infrared excess and might be already a Class II object.

   

  \begin{figure}
     \centering \resizebox{\hsize}{!}{\includegraphics{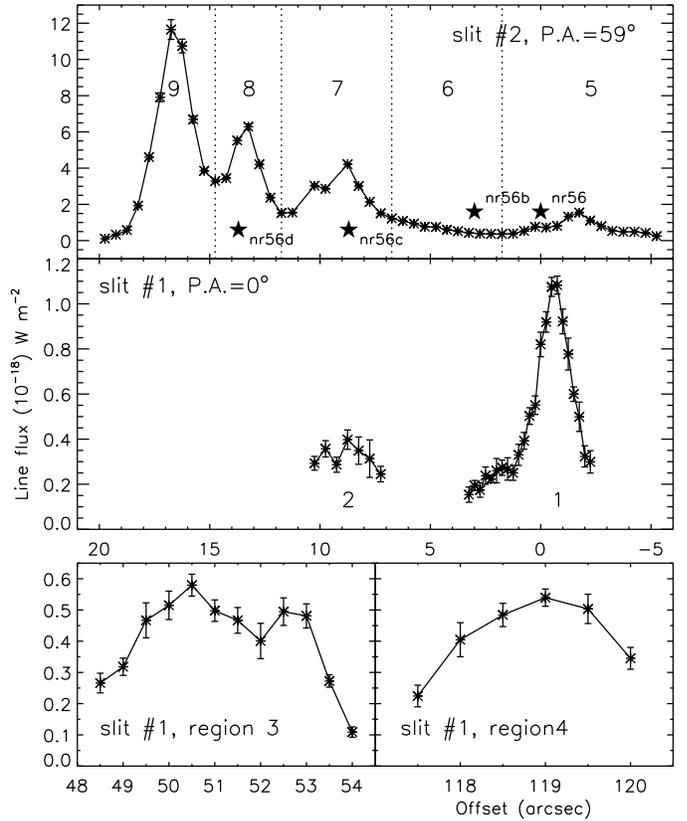}}
     \caption{Spatial variation of the H$_2$ 1-0 S(1) transition at
      2.121~\um\ across slits \#1 and \#2. The origin corresponds to the (0\arcsec, 0\arcsec) position of 
      Fig.~\ref{fig:slits}. The different regions 
      analysed in the text (1 to 9) are identified. Regions 3 and 4 correspond to \Hm\ knots 1 and 2, respectively. The stars indicate the positions of sources nr56, nr56b, nr56c and nr56d.}
     \label{fig:var}
  \end{figure}
  
   \begin{figure*}
    \begin{minipage}[c]{0.5\textwidth}
      \centering \resizebox{\hsize}{!}{\includegraphics{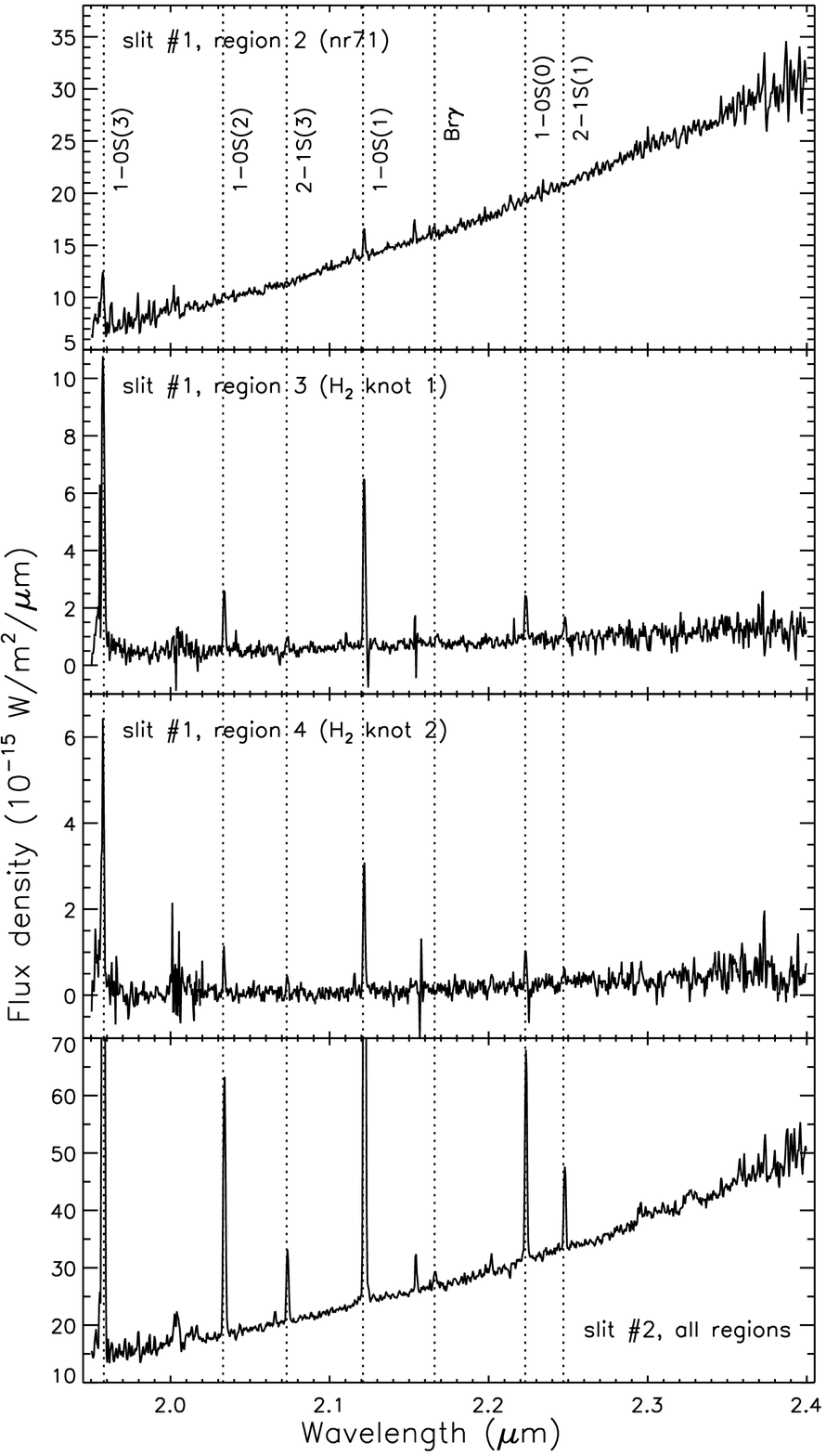}}
    \end{minipage}%
    \begin{minipage}[c]{0.5\textwidth}
      \centering \resizebox{\hsize}{!}{\includegraphics{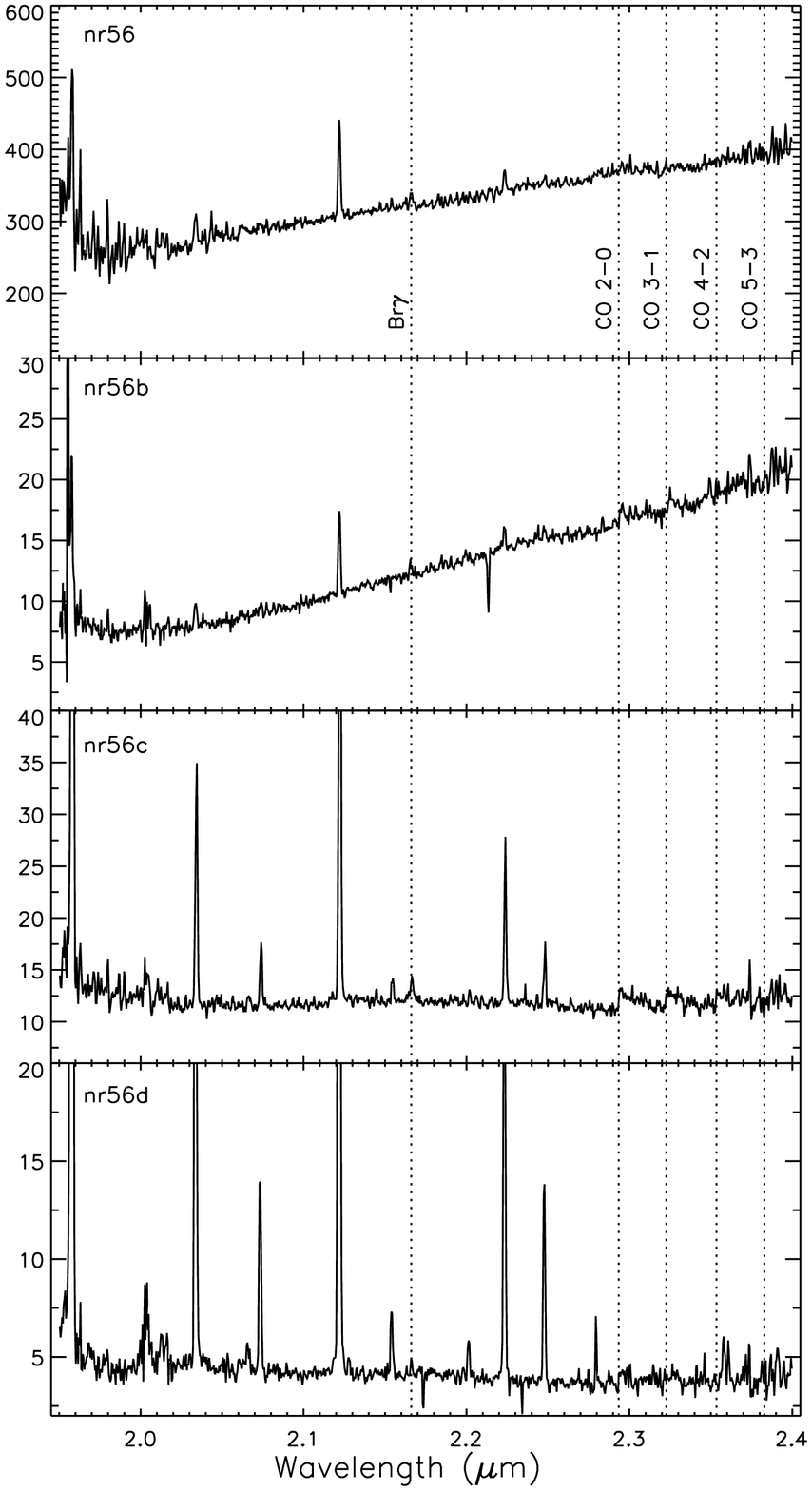}}
    \end{minipage}
       \caption{ {\it (Left)} Spectra of various regions along slit \#1 (P.A.$=0\degr$). The lower panel show the integrated spectra across slit \#2 (P.A.$=59\degr$). The positions of the strongest H$_2$ lines and Br$\gamma$ are indicated by dotted lines.
       {\it (Right)} Spectra of sources nr56, nr56b, nr56c and nr56d. The positions of Br$\gamma$ and the CO bandheads are marked. These four spectra have been corrected for extinction using the values listed in Table~\ref{table:h2}.}
          \label{fig:sp}
  \end{figure*}

\setcounter{table}{1}
\begin{table*}
\caption{Line fluxes measured along slit \#1 (P.A.= $0\degr$) and slit \#2 (P.A.=$50\degr$) in units of $10^{-18}$~W~m$^{-2}$. We also give significant line flux ratios and results
such as the ortho-to-para ratios ($\phi_1$, $\phi_2$ and $\phi_3$), the extinction ($A_K$) and the excitation temperature ($T_{\rm ex}$).}
\label{table:h2}
\begin{center}
\begin{tabular}{l@{\hspace{4pt}}r@{\hspace{8pt}}r@{\hspace{8pt}}r@{\hspace{8pt}}r@{\hspace{8pt}}c@{\hspace{8pt}}r@{\hspace{8pt}}r@{\hspace{8pt}}r@{\hspace{8pt}}r@{\hspace{8pt}}r@{\hspace{8pt}}r}
\hline\hline\\[-5pt]
& \multicolumn{4}{c}{Slit \#1: P.A.$=0\degr$} & & \multicolumn{6}{c}{Slit \#2: P.A.$=59\degr$}\\[1pt]
\cline{2-5} \cline{7-12} \\[-5pt]
           & Region 1 & Region 2 & Region 3 & Region 4 & & Region 5 & Region 6 & Region 7 & Region 8 & Region 9 & 5+6+7+8+9\\[2pt]
\hline\\[-5pt]
1-0 S(3)$^{\rm a}$   & $7.4\pm1.6$  & $10.1\pm2.2$ & $22.1\pm5.9$ & $10.8\pm3.7$ & & $20.1\pm1.6$ & $14.7\pm2.5$ & $75.3\pm4.7$ & $55.2\pm4.4$ & $169\pm42$ & $343\pm22$\\
1-0 S(2)   & $3.0\pm1.2$  & $<4$         & $4.1\pm0.9$  & $1.6\pm0.2$ & & $5.7\pm1.0$  & $4.3\pm0.5$  & $16.0\pm0.5$ & $16.2\pm0.6$ & $38.7\pm1.6$ & $81.8\pm1.9$\\
3-2 S(5)   & $<5$         & $<4$         & $<1$         & $<1$        & & $<3$         & $<2$         & $0.7\pm0.2$  & $0.6\pm0.2$  & $1.6\pm0.3$  & $2.9\pm0.9$ \\
2-1 S(3)   & $<4$         & $<4$         & $1.2\pm0.3$  & $0.6\pm0.3$ & & $1.5\pm0.7$  & $1.5\pm0.3$  & $4.4\pm0.2$  & $3.7\pm0.2$  & $10.8\pm0.5$ & $22.0\pm1.1$\\
1-0 S(1)   & $10.8\pm1.2$ & $4.0\pm0.9$  & $10.2\pm0.9$ & $4.4\pm1.4$ & & $20.5\pm1.0$ & $12.3\pm0.5$ & $47.0\pm0.8$ & $47.0\pm1.5$ & $106\pm5$    & $237\pm5$   \\
3-2 S(4)   & $<4$         & $<4$         & $<1$         & $<1$        & & $<3$         & $<1$         & $<1$         & $<1$         & $0.9\pm0.2$  & $1.3\pm0.9$ \\
2-1 S(2)   & $1.4\pm0.9$  & $2.7\pm0.7$  & $<2$         & $<2$        & & $<5$         & $<3$         & $1.9\pm0.3$  & $1.1\pm0.2$  & $4.8\pm0.3$  & $7.8\pm1.7$ \\
Br$\gamma$ & $<3$         & $<3$         & $<1$         & $<1$        & & $1.6\pm0.7$  & $1.2\pm0.4$  & $1.6\pm0.4$  & $<1$         & $<1$         & $5.4\pm1.7$ \\
3-2 S(3)   & $<6$         & $<6$         & $<1$         & $<1$        & & $<4$         & $<2$         & $<1$         & $0.5\pm0.1$  & $2.5\pm0.2$  & $4.3\pm1.4$ \\
1-0 S(0)   & $2.5\pm1.6$  & $<6$         & $3.0\pm0.4$  & $1.6\pm0.7$ & & $6.4\pm1.4$  & $3.6\pm1.0$  & $11.7\pm0.4$ & $12.0\pm0.5$ & $26.8\pm1.4$ & $60.6\pm2.7$\\
2-1 S(1)   & $<6$         & $<6$         & $1.8\pm0.5$  & $<1$        & & $2.4\pm1.0$  & $1.4\pm0.5$  & $4.4\pm0.4$  & $3.8\pm0.2$  & $11.6\pm0.7$ & $23.8\pm1.8$\\
2-1 S(0)   & $<7$         & $<6$         & $<2$         & $<2$        & & $<5$         & $<3$         & $<3$         & $0.6\pm0.3$  & $2.9\pm0.6$  & $5.1\pm2.6$\\
3-2 S(1)   & $<7$         & $<11$        & $<2$         & $<2$        & & $<8$         & $<6$         & $<4$         & $1.2\pm0.5$  & $<3$         & $7.5\pm4.1$\\[2pt]
\hline\\[-5pt]

${{\rm 1-0~S(1)}\over {\rm 2-1~S(1)}}$\,$^{\rm (b)}$ & $>1.8$       & $>0.7$ & $5.7\pm1.6$  & $>4.4$       & & $8.5\pm3.6$ & $8.8\pm3.1$ & $10.7\pm1.0$ & $12.4\pm0.8$ & $9.1\pm0.7$ & $10.0\pm0.8$\\
${{\rm 1-0~S(1)}\over {\rm 3-2~S(3)}}$\,$^{\rm (b)}$ & $>1.8$       & $>0.7$ & $>10.2$      & $>4.4$       & & $>5.0$      & $>6.1$      & $>47$        & $94\pm19$    & $42\pm4$    & $55\pm18$\\
$\phi_1$          & $3.2\pm1.3$  & $>0.6$ & $2.4\pm0.3$  & $2.2\pm0.8$  & & $2.8\pm0.4$ & $2.6\pm0.4$ & $2.9\pm0.1$  & $2.8\pm0.1$  & $2.8\pm0.1$ & $2.9\pm0.1$\\
$\phi_2$          & $<4.1$       & $<1.9$ & $>1.1$       &              & & $>0.4$      & $>0.6$      & $2.5\pm0.3$  & $3.8\pm0.4$  & $2.6\pm0.1$ & $3.3\pm0.5$\\
$\phi_3$          &              &        &              &              & &             &             &              & $>1.0$       & $2.5\pm0.5$ & $2.8\pm1.5$\\
$A_K$             & $4.5\pm0.4$  &        & $2.8\pm0.5$  & $3.4\pm0.6$  & & $4.0\pm0.2$ & $1.6\pm0.4$ & $1.5\pm0.5$  & $1.2\pm0.7$  & $1.3\pm1.1$ & $1.6\pm0.5$\\
$T_{\rm ex}$ (K)  & $3440\pm265$ &        & $2300\pm115$ & $2295\pm165$ & & $2000\pm40$ & $2270\pm80$ & $2115\pm70$  & $2000\pm70$  & $2390\pm110$ & $2135\pm60$\\[2pt]

\hline
\end{tabular}
\end{center}
$^{\rm (a)}$ This line falls in a region of poor atmospheric transmission; the indentification as 1-0 S(3) is therefore dubious and it is discarded from the analysis.
$^{\rm (b)}$ This line ratio has not been corrected for extinction.
\end{table*}

\subsection{Spatially resolved NIR spectroscopy}

Figure~\ref{fig:var} illustrates the spatial variation of the \Hm\ 1--0 S(1) transition at 2.121~\um. Moving along slit \#1 and from south to north, \Hm\ emission is detected at the position of objects nr56 and nr71 (regions 1 and 2), and about 50\arcsec\ and 119\arcsec\ north of object nr56 (regions 3 and 4, also denominated \Hm\ knots 1 and 2, respectively).
Across slit \#2, the \Hm\ emission extends over $\sim25\arcsec$. We have distinguished 5 different regions in slit \#2, named 5 to 9. Moving from NE to SW, the \Hm\ line emission is stronger in region 9. After this region, the slit crosses over two bright blobs (which correspond to sources nr56d and nr56c, see below), and decreases smoothly afterwards.

The left panel of Fig.~\ref{fig:sp} shows the nebular spectra of \iras\ extracted from regions 2, 3 and 4 of slit \#1, and the integrated spectrum across slit \#2. The most obvious features of these spectra are a large number of strong molecular hydrogen transitions and a weak Br$\gamma$ emission line. Table~\ref{table:h2} lists the identified lines, their central wavelengths and fluxes. The quoted uncertainties are from the line fitting procedure. 

We also extracted four spectra along slit \#2 where an unresolved underlying continuum was detected. In particular, at 0\arcsec\ (which coincides with object nr56), at 3\arcsec\ (which we will denominate hereafter nr56b), at 8\farcs7 (nr56c) and at 13\farcs7 (nr56d). Since no photospheric features are detected, origins for the continua other than stellar must be considered. Moreover, a stellar origin is unlikely because of the relatively large extinction towards \iras\ (see Sect.~\ref{section:h2}). These continua can also be produced by (1) free-free and free-bound emission from ionized gas, (2) scattered light and (3) hot dust. The free-free continuum emission detected at radio wavelengths contributes only marginally to the infrared continuum detected in the $K-$band. This was estimated using equation 7 of \cite{roman04} and the free-free Gaunt factors of \cite{hummer88}. In particular, for a flux density of 1~mJy at 3.6~cm and a typical electron temperature of $10^4$~K, the expected free-free flux density in the $K-$band is only about 0.15~mJy, which translates into $\sim10^{-16}$~W~m$^{-2}$~\um$^{-1}$ at 2.2~\um, less than about 1\% of the observed continua. Most likely, the continua are caused by hot dust as the spectra of e.g. nr56 and nr56b are very red. These red slopes cannot be explained by scattered light, whose effect becomes stronger towards shorter wavelengths.

The spectra of nr56, nr56b, nr56c and nr56d are illustrated in the right panel of Fig.~\ref{fig:sp}. The positions of these sources within \iras\ are indicated in Fig.~\ref{fig:slits}. Besides the nebular \Hm\ lines and weak Br$\gamma$, an important result is the detection of CO first-overtone bandhead emission lines in nr56b and nr56c. The spectrum of nr56d has a very faint continuum and is mostly dominated by \Hm\ lines. These four spectra have been corrected for extinction using the values listed in Table~\ref{table:h2}, i.e. $A_K=4.0$~mag for nr56 and $A_K=1.6$~mag for the other 
sources. The derivation of these values will be explained later in Sect.~\ref{section:h2}. It is interesting to note that the continuum slopes of sources nr56c and n56d are not very red after dereddening. The red continuum slopes of sources nr56 and nr56b, as mentioned previously, can be easily explained by emission of hot dust.

\subsection{\Hm\ excitation}
\label{section:h2}

The NIR emission from H$_2$ can be produced by either thermal emission in shock
fronts or fluorescence excitation by non-ionizing UV photons in the Lyman-Werner band (912--1108 \AA). These mechanisms can be distinguished since they preferentially populate different levels producing different line ratios. 

Non-thermal excitation mechanims readily excite the $v=2$ and higher vibrational states, whereby collisional transitions preferentially de-excite the $v=2$ level in favor of the $v=1$ level. Hence, the H$_2$ 1--0 S(1)/2--1 S(1) ratio has on occasion been used to distinguish between these two processes, with a pure fluorescense value around 2 and a thermal value of order 10 \citep[e.g.][]{davies03}. However, for high density clouds ($n_{\rm H} \gtrsim 10^5$ cm$^{-3}$) and for an intense incident FUV radiation field with $G_0 \gtrsim 10^4$ (where $G_0$ is the incident FUV photon flux between 6 and 13.6~eV measured in units of the local interstellar radiation field), collisional de-excitation of FUV-pumped molecules will thermalize the lower energy levels and the 1--0 S(1)/2--1 S(1) ratio will rapidly approach a thermal value \citep[e.g.][]{burton90,hollenbach95,davies03}. Conversely, the higher vibrational states are expected to retain a fluorescent population. For instance, the 1--0 S(1)/3--2 S(3) ratio retains a value of approximately 8, and only for very high densities and a very intense FUV radiation field will this ratio approach a thermal value of 10--100. Therefore, the combination of the 1--0 S(1)/2--1 S(1) and 1--0 S(1)/3--2 S(3) ratios should indicate the H$_2$ excitation mechanism.

Table~\ref{table:h2} lists the 1--0 S(1)/2--1 S(1) and 1--0 S(1)/3--2 S(3) ratios across slit \#1 (P.A.$=0\degr$) and slit \#2 (P.A.$=59\degr$). Across slit \#2, the 1--0 S(1)/2--1 S(1) ratio remains practically constant with an average value around 10. The 1--0 S(1)/3--2 S(3) ratio, on the other hand, has an average value of 55 and reaches a value as high as 100 in region 8. The combination of both ratios suggests the presence of shocks across this bulk of emission. This analysis is however not conclusive in the case of slit \#1, even though the dense PDR models of \cite{davies03} seem to match the ratios observed for region 3.

An additional diagnostic is provided by the ortho-to-para ratio. At the moment of formation, molecular hydrogen can join one of two denominations: ortho (aligned nuclear spins) or para (opposed nuclear spins). The ortho-to-para ratio $\phi$, based on the statistical weights of the nuclear spins, is 3:1. 
In outflow regions where shock excitation is the primary emission mechanism, observations of vibrational emission lines typically reveal ortho-to-para ratios for vibrationally excited states of about 3 \citep[e.g.][]{smith97}.
However, as discussed at length by \cite{sternberg99}, the ortho-to-para ratio can take values other than the intrinsic 3 if the excitation is caused by UV fluorescence. Indeed, ratios in the range 1.5--2.2 have been measured in PDRs \citep[e.g.][]{ramsay93,chrysostomou93,shupe98,lumsden01}.

From the 1--0 S(0), 1--0 S(1) and 1--0 S(2) line fluxes we measure the ortho-to-para ratio for the $v=1$ states ($\phi_1$) following equation 9 of \cite{smith97}. In Table~\ref{table:h2} we list the resulting $\phi_1$ for each region. In general, these are consistent with a ratio of 3 suggesting a thermalization of the $v=1$ states. 

  \begin{figure*}
     \centering \resizebox{\hsize}{!}{\includegraphics{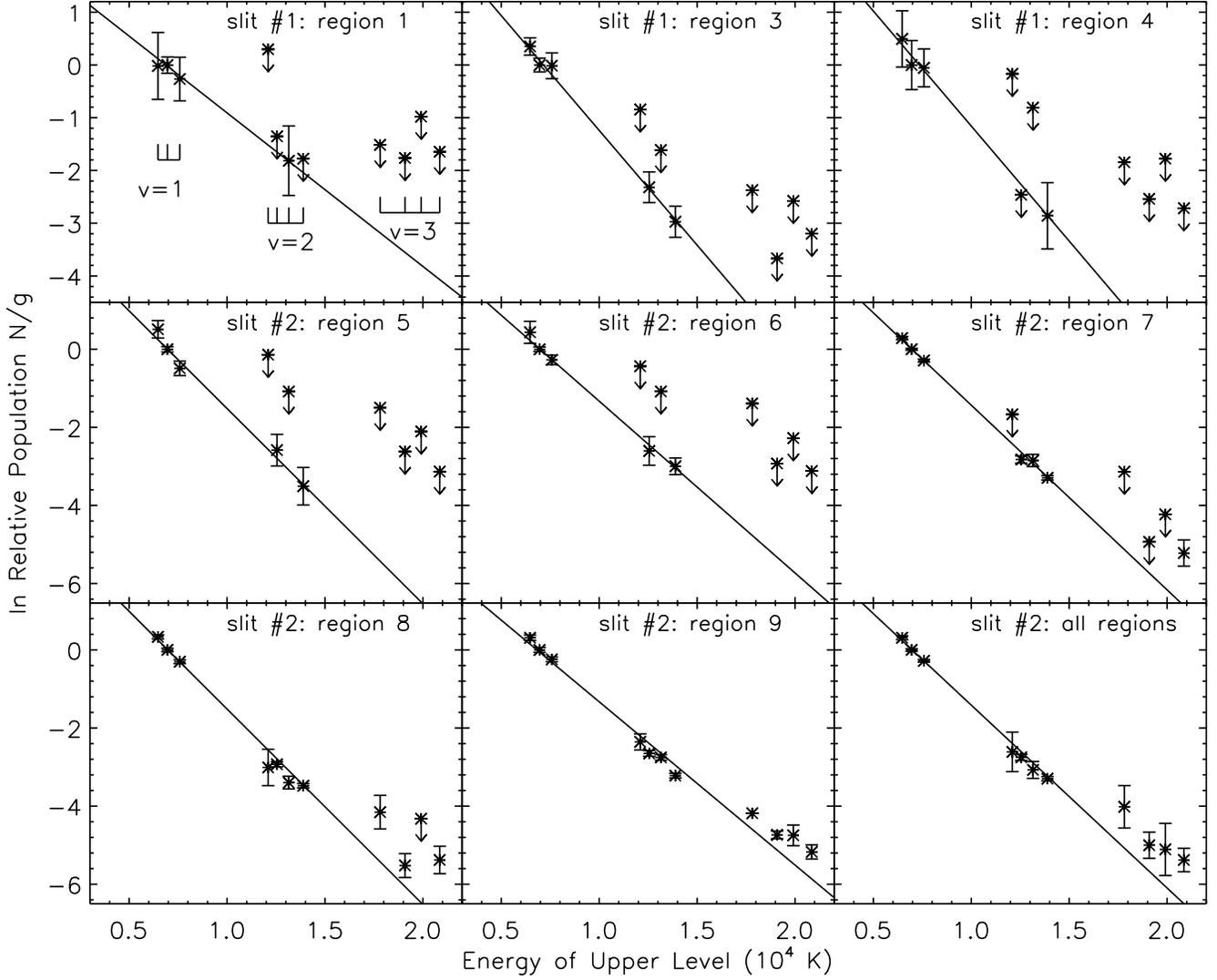}}
     \caption{H$_2$ excitation diagrams for every region (except region 2, where only two useful lines are available). The lines represent the best fitting purely thermal single-temperature models.}
     \label{fig:h2}
  \end{figure*}

Of course, $\phi_1$ does not tell the whole story. A value of the ortho-to-para ratio for the $v=2$ states ($\phi_2$) is derived from the 2--1 S(1), 2--1 S(2) and 2--1 S(3) transitions using equation 1 of \cite{davis03}.  Although the values for $\phi_2$ are less certain because the $v=2$ transitions are in general weak, the results in Table~\ref{table:h2} indicate that this ratio is in general close to 3. This suggests a thermalization of the $v=2$ levels, although we find $\phi_2<1.9$ in region 2, a value which is typical of PDRs. 

In a few cases we could derive a value for the $v=3$ levels using the 3--2 S(3), 3--2 S(4) and 3--2 S(5) transitions. Using the methodology of \cite{smith97}, the ratio from these transitions is given by:

\begin{equation}
1/\phi_3 = 0.945(I_4/I_5)^{0.467}(I_4/I_3)^{0.533}~. 
\end{equation}

The values we obtain for slit \#2 are again close to 3 as can be seen in Table~\ref{table:h2}.

A more sophisticated way of characterizing the \Hm\ emission is by plotting the observed column density against the energy of the upper level. The measured intensity, $I$, of a given H$_2$ line can be used to calculate the column density of the upper excitation level of the transition, which, for optically thin emission, is given by:

\begin{equation}
 N_j = { {4 \pi \lambda_j I} \over {A_j h c} }~,
\end{equation}

\noindent
where $\lambda_j$ is the rest wavelength and $A_j$ is the Einstein A-coefficient taken from \cite{wolniewicz98}. If collisional de-excitation is assumed to dominate, the \Hm\ will be in LTE and the energy levels can be described by a Boltzmann distribution. The relative column densities of any two excitation levels can thus be expressed in terms of an excitation temperature $T_{\rm ex}$:

\begin{equation}
 { N_i \over N_j} = {g_i \over g_j} {\rm exp} \left [ { {-(E_i -E_j)} \over {k T_{\rm ex}} } \right ]~,
\end{equation}

\noindent
where $g_j$ is the degeneracy, $E_j$ is the energy of the upper level taken from \cite{dabrowski84} and $k$ is the Boltzmann's constant. The values of $\lambda_j$, $E_j$, $A_j$ and $g_j$ for the lines detected in the spectra are given in Table~\ref{table:h2:theo}. Clearly, plotting the logarithm of the ratio of the column densities divided by the appropriate statistical weights gives the excitation temperature from the slope of the data.

In Fig.~\ref{fig:h2} we show plots of log$N_j/g_j$ versus the energy. We have normalized the population distributions relative to that inferred from the 1--0 S(1) line and corrected for extinction. A standard extinction law of the form $A_\lambda \propto \lambda^{-1.7}$ \citep[e.g.][]{mathis90,martin90} was used. For thermalized populations at a fixed gas temperature, the log$N/g$ points should lie on a straight line in these excitation diagrams. If the value of $A_k$ used to correct the line intensities from which the column density is calculated is wrong, then the scatter of the points will increase. The values of $A_k$ that minimize $\chi^2$ and the resulting excitation temperatures are reported in Table~\ref{table:h2}. Of course, these values of the extinction are estimated assuming a purely thermal single-temperature gas and thus, they must be taken with caution. Temperatures range from 2000 to 3400~K, while the extinction spreads from approximately $A_k=1$~mag to $A_k=4.5$~mag.

For fluorescence emission, each vibrational level has $N_j/g_j$ lying along a separate ``branch'' and the rotational population within each level can be approximated by a thermal distribution. A curved line in this plot would therefore provide evidence for non-LTE processes. In the case of \iras,
a single temperature LTE gas fits rather well the $v=1$ and $v=2$ transitions indicating that at least these two levels appear to be thermalized at temperatures of the order of 2000~K. Unfortunately, this cannot help to discriminate between shocks and UV excitation. This is because in dense PDRs the lower \Hm\ levels will be thermalized as in shocks. However, the distinction can be made by looking at the $v=3$ transitions, which will be predominantly excited by the PDR.

When $v=3$ transitions are detected (as is the case of slit \#2), their column densities are only slightly underpredicted by the single component models. This suggests that a non-thermal mechanism might be partially responsible for the excitation of these high vibrational levels.
However, the fact that 1--0 S(1)/\Bg\ is $\ggg$1 strongly argues against a stellar UV excitation model \citep[cf.][]{hatch05}. Moreover, the faintness of \Bg\ emission and the measured excitation temperatures suggest that fast C-shocks might be present in \iras. This type of shock has typical velocities of the order of 40 km\,s$^{-1}$ and heats the gas to $\sim 2000$~K producing strong \Hm\ lines \citep[e.g.][]{davies00}. The UV radiation field necessary to excite the $v=3$ levels of \Hm\ can be produced in the dissociated apex of a bow-shaped shock \citep{fernandes95,fernandes97}. The excess emission observed for the lines with $v=3$ would then arise from \Hm\ fluorescence produced by Ly$\alpha$ pumping of the low density ($\sim 10^3-10^4$ cm$^{-3}$) pre-shocked gas, while the bulk \Hm\ emission is excited in a bow C-shock. 

\subsection{The origin of the Br$\gamma$ emission}
\label{section:bg}

Recombination radiation from hydrogen atoms is observed through the Br$\gamma$ transition ($\lambda=2.166$~\um) in regions 5, 6 and 7 (see Table~\ref{table:h2}), and is present in the spectra of nr56, nr56b and nr56c. 

\cite{beuther02b} reported the detection of a weak \mbox{($\sim 1$\,mJy)} radio continuum source towards nr56. Using equation 8 of \cite{martin:atca:gal} and considering optically thin emission, a dust-free \HII\ region with such 3.6~cm emission will produce a \Bg\ line flux of about $1\times10^{-17}$ W\,m$^{-2}$. Reddening this line flux (we use an extinction $A_K=4$, see Table~\ref{table:h2}), this translates into an observed \Bg\ line flux of $\sim 2.5\times10^{-19}$ W\,m$^{-2}$, i.e. about 15\% of the \Bg\ line flux measured in region 5. Other origins might be necessary to explain the \Bg\ line emission detected in this region. In fact, \Bg\ emission can also arise in shocks. It has been detected, for instance, in the bases of jets from T Tauri stars and in the accretion flows around low mass Class I sources \cite[e.g.][]{folha01,davis01}.


The Br$\gamma$ emission detected in the spectra of nr56b (region 6) and nr56c (region 7), however, probably have a different origin. 
The detection of CO bandheads in these spectra suggests the presence of dense circumstellar material, possibly a disk, implying that the Br$\gamma$ emission could have the same origin. It could be produced, for instance, in a disk wind caused by the interaction between the UV photons and the ionized upper layer of the disk.

\begin{table}
\caption{We give the air wavelengths, upper energy levels, Eistein A 
coefficients and degeneracies of the H$_2$ lines detected in \iras.}
\label{table:h2:theo}
\begin{center}
\begin{tabular}{ccccc}
\hline\hline\\[-5pt]
Transition & $\lambda_{\rm j}$ & $E_{\rm j}$ & $A_{\rm j}$ & $g_{\rm j}$\\[1pt]
           & (\um) & (K) & (10$^{-7}$ s)\\[2pt]
\hline\\[-5pt]

1-0 S(3) & 1.9571 &  8365 & 4.21 & 33\\
1-0 S(2) & 2.0332 &  7584 & 3.98 &  9\\
3-2 S(5) & 2.0650 & 20856 & 4.53 & 45\\
2-1 S(3) & 2.0729 & 13890 & 5.77 & 33\\
1-0 S(1) & 2.1212 &  6956 & 3.47 & 21\\
3-2 S(4) & 2.1274 & 19912 & 5.25 & 13\\
2-1 S(2) & 2.1536 & 13152 & 5.60 &  9\\
3-2 S(3) & 2.2008 & 19086 & 5.63 & 33\\
1-0 S(0) & 2.2229 &  6471 & 2.53 &  5\\
2-1 S(1) & 2.2471 & 12550 & 4.98 & 21\\

2-1 S(0) & 2.3549 & 12095 & 3.68 & 5\\
3-2 S(1) & 2.3857 & 17818 & 5.15 & 21\\[2pt]
\hline
\end{tabular}
\end{center}
\end{table}

\section{Discussion and conclusions}
\label{sect:conclusions}
 
The new NIR spectroscopic data presented here has provided, on the one hand, a complete analysis of the \Hm\ emission and, on the other hand, direct evidence for the presence of massive YSOs and their circumstellar environment. 

\subsection{Stellar content}
\label{sect:con:part1}

The NIR emission of \iras\ mainly concentrates on an elongated feature suffering an extinction $A_V\sim 15$ mag. This structure is aligned with the blue-shifted molecular emission of a CO outflow \citep{beuther03}. At the emission maximum of this almost linear structure, where the source nr56 is located, the extinction is even higher, with $A_V\sim 40$ mag. This source nr56 is not only associated with the brightest MIR source, but also with the millimeter source mm1, which has a mass of a few tens of solar masses \citep{beuther04}. 
A spherical distribution of this cold material would lead to a very large visual extinction ($A_V\sim1000$). No emission at near-infrared, or even at mid-infrared wavelengths, would thus be expected. However, the observation of NIR and MIR counterparts show that emission is leaking at these wavelengths. One explanation would be that the NIR and MIR emission is coming through a cavity created by one of the many outflows present in this region. The NIR and MIR emission could be emitted in the close environment of the forming stars hidden in mm1 and escape via the outflow cavity. Another possibility is that the IR emission is coming from the outflow cavity itself, i.e. from dust heated on the walls of the outflow cavity near the source \citep{deBuizer05}.
In this picture, the mm emission would be associated with the outer regions of the circumstellar matter.

  \begin{figure}
     \centering \resizebox{\hsize}{!}{\includegraphics{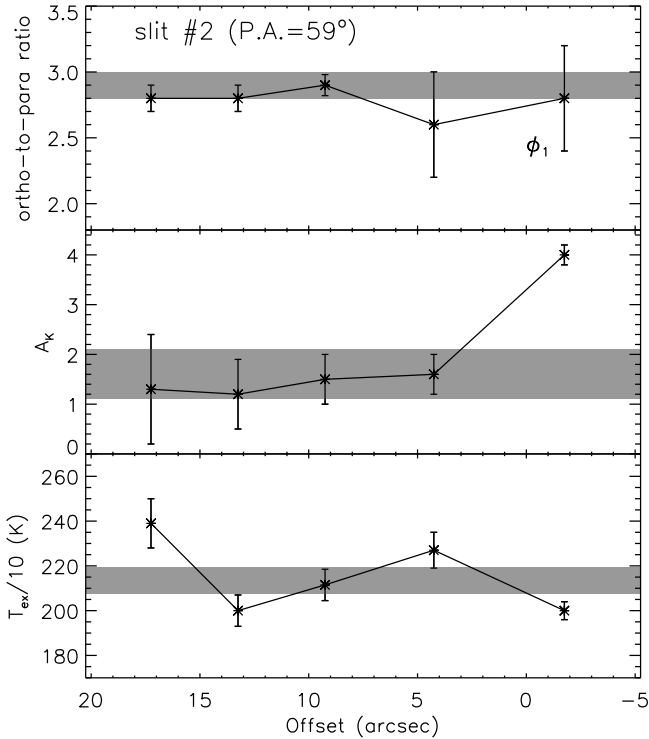}}
     \caption{Spatial variation across slit \#2 (P.A.$=59\degr$) of the ortho-to-para ratio ($\phi_1$), the extinction ($A_{\rm K}$) and the excitation temperature ($T_{\rm ex}$). In grey we show the average results from the integrated spectrum across this slit. Offsets are the same as in the top panel of Fig.~\ref{fig:var}.}
     \label{fig:h2:results}
  \end{figure}

Source nr56 also exhibits a large [3.6]--[4.5] excess, which indicates that it might be an embedded Class 0/I YSO. However, the detection of 3.6\,cm continuum emission suggests in principle the presence of a recently ignited, and thus a more evolved star which has already formed an ultracompact \HII\ region. 
Assuming for now that there is indeed an Ultracompact \HII\ region --it could probably 
be a Hypercompact \HII\ region \citep{kurtz05}-- and that the radio emission is optically thin, it would be consistent with an ionizing B2\,V star \citep{panagia73}. However, in the case of the high-mass protostar NGC\,7538\,IRS1, \cite{deBuizer05} note that the centimeter continuum emission detected in this object could be the partially ionized emission from an outflow, or even, could arise from a photoevaporated disk wind.
Shocks associated with the supersonic infall of matter onto a protostellar disk are also sources of radio free-free emission. This last scenario can however be excluded since the models predict fluxes much lower than 1~mJy \citep{neufeld96}. 
Radio observations of the contiuum emission at other frequencies will be most helpful to determine the nature of this emission, i.e. whether it is optically thin or optically thick.
Another clue to the nature of nr56 is the detection of H$_2$O and \ion{Class}{ii} CH$_3$OH masers \citep{beuther02b}, which indicates a very early evolutionary stage. Both maser types, although they are on occasion spatially coincident with ultracompact \HII\ regions, are  more often found in hot molecular cores and/or massive molecular outflows. In particular, H$_2$O masers are believed to be excited by collisional pumping with \Hm\ molecules within shocks associated with outflows and/or accretion \citep[e.g.][]{elitzur89,garay99}. Finally, its $K-$band spectrum only shows weak Br$\gamma$ and is practically featureless. It has a red continuum slope which is probably caused by hot dust. 

Sources nr56b and nr56c, detected towards the east of nr56, show weak Br$\gamma$ and CO first-overtone bandhead emission in their spectra. This is the first detection of CO bandheads in \iras. These CO bandheads are observed in about 25\% of all the massive YSOs \citep[e.g.][]{chandler93,chandler95,bik04,blum04} and are emitted in neutral material with temperatures between 2000 and 5000~K, and densities of about $10^{10}$~cm$^{-3}$. This hot neutral gas is probably located in the inner, dust-free regions of the circumstellar environment, relatively close to the star. \cite{bik04} modeled the profiles of the CO bandheads in 4 massive YSOs and suggested that they could be explained by a simple Keplerian disk, probably the remnants of massive accretion disks.
Source nr56b also has a weak millimeter counterpart \citep{beuther04}. Both sources appear to be less embedded --and perhaps more evolved-- YSOs than nr56. 
Source nr56d, on the other hand, with a spectrum dominated by \Hm\ lines, is presumably a knot or bow shock in the outflow.

Another source of interest is nr71, located 9\arcsec\ north of nr56. It has not a large IR excess and might be already a Class II object. It has a featureless $K-$band spectrum. The lack of Br$\gamma$ emission could indicate that the circumstellar material is not (or very midly) ionized.

Finally, the millimeter source mm2, located 4\arcsec\ north of nr56, is associated with MIR emission only visible at $\lambda \gtrsim 6$~\um. This source, which is nearly as massive as mm1 \citep{beuther04}, could be a candidate for a high-mass protostar, the equivalent of Class O low mass YSOs which are still in the main accretion phase. 

Other sources in the vicinity of \iras\ show infrared excess at wavelengths longward of about 3~\um. In particular, sources 2MASS\_1, nr14, nr34 and nr64 have a large MIR excess, suggesting that they might be embedded YSOs. 2MASS\_1, nr14 and nr34 are associated with X-ray emission. Interestingly, there are not many objects around \iras\ with a NIR excess. 

\subsection{The nature of the H$_2$ emission}

Based on the \Hm\ line ratios, the measured ortho-to-para ratios and results of the excitation diagrams, we conclude that the \Hm\ emission observed across slit \#2 is clearly produced by shocks.  Of course, the fact that the linear emission feature mapped by this slit is aligned with one of the outflows found by \cite{beuther03} reinforces the shocked nature of this \Hm\ emission.

Figure~\ref{fig:h2:results} shows the spatial variation of the excitation temperature and extinction across slit \#2. The temperature is practically constant with a value around 2100~K, increasing slightly up to 2400~K in region 9. The extinction $A_k$, on the other hand, is higher in region 5, where it reachs a value of 4.0, and keeps a lower and almost constant value around 1.5 from regions 6 to 9.  

The conclusions for the regions observed across slit \#1 are more ambiguous because of the lower number of \Hm\ lines detected. It is thus difficult to discriminate between shocks and dense PDRs. The low values found for $\phi_1$ and $\phi_2$ in regions 2, 3 and 4 seem to favor the case of dense PDRs. 
However, the presence of shocks cannot be excluded. \citeauthor{neufeld98} (\citeyear{neufeld98}, \citeyear{neufeld06}) have measured ortho-to-para ratios in Herbig-Haro objects and found values significantly smaller than the equilibrium ratio of 3. They conclude that these low ratios imply that the gas has not been warm long enough to reach equilibrium between the ortho and para states and that they are the legacy of an earlier stage in the thermal history of the gas when the gas reached equilibrium at a temperature $<90$~K. They even place a conservative upper limit of $\sim 5000$~yr on the period on which the emitting gas has been warm. This upper limit is based upon the expected timescale for the conversion from para- to ortho-\Hm.
Hence, the low ortho-to-para ratios we measured in these three regions could indicate either the presence of a very recent shocked gas or of a dense PDR.

\subsection{Conclusions}

We obtained long slit, intermediate-resolution, NIR spectra of \iras\ along two preferential directions, i.e. P.A.=0\degr\ and P.A.=59\degr. As a complement, we also obtained $J$, $H$ and $K_s$ images taken with the Las Campanas Du Pont Telescope, which we combined with archival \spitzer\ data.

We have found objects with very different properties and evolutionary stages in \iras. In particular, the NIR spectra show CO first-overtone bandhead emission which is associated with neutral material located in the inner regions of the circumstellar environment of YSOs. The brightest source is nr56, which is probably the driving source of the strongest outflow. It is also the most massive source at millimeter wavelengths, with a mass of a few tens of solar masses, and its early evolutionary stage is confirmed by the presence of H$_2$O and CH$_3$OH masers. The fact that nr56 is detected in the near- and mid-infrared suggests that the assumption of a spherical distribution is probably not right. Such a spherical distribution of cold material would lead to a visual extinction of up to 1000~magnitudes.
Emission is likely leaking at these wavelengths, coming through a cavity created by one of the outflows present in the region, or from the outflow cavity itself. 

The second most massive millimeter source, mm2, is only detected at 
$\lambda \gtrsim 6$\,\um, suggesting that it could be a high-mass protostar still in its main accretion phase. 

Another important conclusion is the confirmation of the shocked nature of the \Hm\ emission, with an excitation temperature of about 2000~K, based on the analysis of relevant \Hm\ line ratios, ortho-to-para ratios and excitation diagrams. \Hm\ emission at 50\arcsec\ and 119\arcsec\ north of nr56 is also found, but we could not determine whether it originates in shocks or in a dense PDR. If it is produced in shocks, this emission could be associated with the northern millimeter clump, where at least three outflows are present \citep{beuther03}.

\begin{acknowledgements}
We thank the referee, C. J. Davis, for his critical reading and constructive comments.
We also thank Henrik Beuther for a helpful discussion on the nature of the millimeter sources.
During this work, NLMH has been supported by a Juan de la Cierva fellowship from the Spanish Ministerio de Ciencia y Tecnolog\'{\i}a (MCyT). This work has also 
been partially funded by the Spanish MCyT under project AYA2004-07466.
LB acknowledges support by Chilean FONDAP Center of Astrophysics 15010003
\end{acknowledgements}


\Online

\setcounter{table}{0}
\clearpage
\onecolumn
\begin{landscape}
 \begin{longtable}{cccccccccc}

\caption{Photometry towards \iras.} \\
\hline\hline\\[-5pt]
ID & R.A. (J2000) & Dec (J2000) & $J$ & $H$ & $K_s$ & [3.6] & [4.5] & [5.8] & [8.0]\\[2pt]
\hline\\[-5pt]
\endfirsthead

\caption{continued.}\\
\hline\hline\\[-5pt]
ID & R.A. (J2000) & Dec (J2000) & $J$ & $H$ & $K_s$ & [3.6] & [4.5] & [5.8] & [8.0]\\[2pt]
\hline\\[-5pt]
\endhead
\hline
\endfoot

  1 & 19:43:08.92 & 23:43:31.2 & 14.24 $\pm$ 0.11 & 13.79 $\pm$ 0.13 & 
13.58 $\pm$ 0.11 & 13.46 $\pm$ 0.08 & 13.59 $\pm$ 0.15 & ... & ... \\
  2 & 19:43:08.37 & 23:43:31.4 & ... & ... & 15.61 $\pm$ 0.13 & 13.02 $\pm$ 0.07
 & 12.53 $\pm$ 0.13 & 11.80 $\pm$ 0.20 & ... \\
  3 & 19:43:12.46 & 23:43:32.4 & 15.56 $\pm$ 0.12 & 13.23 $\pm$ 0.13 & 
12.03 $\pm$ 0.10 & 11.27 $\pm$ 0.07 & 11.16 $\pm$ 0.08 & 10.56 $\pm$ 0.17 & ...
 \\
  4 & 19:43:09.52 & 23:43:33.1 & 14.51 $\pm$ 0.11 & 14.26 $\pm$ 0.14 & 
13.92 $\pm$ 0.12 & 13.87 $\pm$ 0.07 & 13.90 $\pm$ 0.20 & ... & ... \\
  5 & 19:43:08.29 & 23:43:33.5 & ... & ... & 15.68 $\pm$ 0.14 & ... & ... & ...
 & ... \\
  6 & 19:43:12.63 & 23:43:34.1 & ... & 15.86 $\pm$ 0.17 & 13.95 $\pm$ 0.12 & 
13.04 $\pm$ 0.13 & 12.71 $\pm$ 0.14 & ... & ... \\
  7 & 19:43:07.55 & 23:43:35.4 & 17.55 $\pm$ 0.20 & 16.40 $\pm$ 0.19 & 
15.82 $\pm$ 0.15 & ... & ... & ... & ... \\
  8 & 19:43:13.26 & 23:43:35.6 & ... & 16.49 $\pm$ 0.23 & 15.49 $\pm$ 0.18 & ...
 & ... & ... & ... \\
  9 & 19:43:08.48 & 23:43:36.4 & 17.75 $\pm$ 0.21 & 16.09 $\pm$ 0.19 & 
15.68 $\pm$ 0.15 & ... & ... & ... & ... \\
 10 & 19:43:13.40 & 23:43:36.2 & 16.66 $\pm$ 0.18 & 14.80 $\pm$ 0.14 & 
13.78 $\pm$ 0.11 & 13.03 $\pm$ 0.10 & 12.93 $\pm$ 0.14 & ... & ... \\
 11 & 19:43:13.63 & 23:43:36.3 & ... & ... & 15.33 $\pm$ 0.16 & ... & ... & ...
 & ... \\
 12 & 19:43:11.00 & 23:43:37.0 & ... & 16.17 $\pm$ 0.21 & 14.01 $\pm$ 0.11 & 
12.85 $\pm$ 0.10 & 12.45 $\pm$ 0.11 & ... & ... \\
 13 & 19:43:12.20 & 23:43:36.9 & ... & ... & 15.51 $\pm$ 0.15 & ... & ... & ...
 & ... \\
 14 & 19:43:11.49 & 23:43:38.1 & 14.58 $\pm$ 0.11 & 12.95 $\pm$ 0.12 & 
11.69 $\pm$ 0.09 & 10.64 $\pm$ 0.05 &  9.95 $\pm$ 0.06 &  9.15 $\pm$ 0.07 & 
 8.47 $\pm$ 0.05 \\
 15 & 19:43:07.81 & 23:43:39.1 & ... & ... & 15.65 $\pm$ 0.18 & 14.48 $\pm$ 0.17
 & ... & ... & ... \\
 16 & 19:43:07.52 & 23:43:39.5 & 16.70 $\pm$ 0.16 & 15.41 $\pm$ 0.16 & 
15.24 $\pm$ 0.13 & ... & ... & ... & ... \\
 17 & 19:43:10.72 & 23:43:40.9 & ... & 13.17 $\pm$ 0.13 & 10.96 $\pm$ 0.09 & 
 9.66 $\pm$ 0.08 &  9.40 $\pm$ 0.06 &  9.00 $\pm$ 0.05 &  8.95 $\pm$ 0.08 \\
 18 & 19:43:09.12 & 23:43:41.4 & ... & ... & 16.57 $\pm$ 0.23 & ... & ... & ...
 & ... \\
 19 & 19:43:12.62 & 23:43:41.8 & ... & 16.55 $\pm$ 0.20 & 15.20 $\pm$ 0.13 & 
14.34 $\pm$ 0.14 & ... & ... & ... \\
 20 & 19:43:08.10 & 23:43:42.3 & ... & ... & 14.21 $\pm$ 0.12 & 10.49 $\pm$ 0.07
 &  9.45 $\pm$ 0.05 &  8.54 $\pm$ 0.05 &  7.79 $\pm$ 0.05 \\
 21 & 19:43:11.60 & 23:43:42.8 & 15.57 $\pm$ 0.15 & 14.12 $\pm$ 0.14 & 
12.85 $\pm$ 0.10 & 11.79 $\pm$ 0.07 & 11.12 $\pm$ 0.08 & 10.15 $\pm$ 0.21 & 
 9.46 $\pm$ 0.08 \\
 22 & 19:43:10.53 & 23:43:42.7 & ... & ... & 15.61 $\pm$ 0.23 & ... & ... & ...
 & ... \\
 23 & 19:43:09.26 & 23:43:43.7 & 17.53 $\pm$ 0.35 & 16.77 $\pm$ 0.25 & 
15.77 $\pm$ 0.17 & ... & ... & ... & ... \\
 24 & 19:43:12.41 & 23:43:44.7 & ... & ... & 15.43 $\pm$ 0.19 & ... & ... & ...
 & ... \\
 25 & 19:43:11.46 & 23:43:45.2 & 11.40 $\pm$ 0.08 & 11.23 $\pm$ 0.11 & 
11.09 $\pm$ 0.09 & 11.14 $\pm$ 0.06 & 10.87 $\pm$ 0.13 &  9.93 $\pm$ 0.20 & 
 9.97 $\pm$ 0.16 \\
 26 & 19:43:09.49 & 23:43:46.0 & 17.02 $\pm$ 0.17 & 15.34 $\pm$ 0.15 & 
14.31 $\pm$ 0.12 & 13.22 $\pm$ 0.15 & 13.05 $\pm$ 0.30 & ... & ... \\
 27 & 19:43:12.20 & 23:43:46.5 & ... & ... & 16.79 $\pm$ 0.31 & ... & ... & ...
 & ... \\
 29 & 19:43:11.86 & 23:43:48.7 & ... & 16.17 $\pm$ 0.18 & 14.11 $\pm$ 0.12 & 
13.02 $\pm$ 0.12 & 13.02 $\pm$ 0.34 & 10.75 $\pm$ 0.33 & ... \\
 30 & 19:43:12.36 & 23:43:47.8 & 13.78 $\pm$ 0.11 & 12.99 $\pm$ 0.12 & 
12.56 $\pm$ 0.10 & 12.06 $\pm$ 0.06 & 11.88 $\pm$ 0.13 & ... & ... \\
 31 & 19:43:12.84 & 23:43:51.0 & ... & ... & 15.82 $\pm$ 0.22 & ... & ... & ...
 & ... \\
 33 & 19:43:10.09 & 23:43:53.6 & 15.10 $\pm$ 0.11 & 14.62 $\pm$ 0.14 & 
14.07 $\pm$ 0.12 & ... & ... & ... & ... \\
 34 & 19:43:11.83 & 23:43:54.2 & ... & ... & 11.55 $\pm$ 0.11 &  9.74 $\pm$ 0.05
 &  9.12 $\pm$ 0.11 &  8.52 $\pm$ 0.05 &  7.91 $\pm$ 0.05 \\
 35 & 19:43:08.80 & 23:43:54.9 & ... & ... & 16.71 $\pm$ 0.21 & ... & ... & ...
 & ... \\
 36 & 19:43:13.48 & 23:43:56.0 & ... & ... & 15.17 $\pm$ 0.14 & ... & ... & ...
 & ... \\
 37 & 19:43:09.49 & 23:43:55.8 & 16.30 $\pm$ 0.16 & 15.54 $\pm$ 0.16 & 
14.94 $\pm$ 0.13 & ... & ... & ... & ... \\
 38 & 19:43:07.72 & 23:43:55.6 & ... & ... & 14.77 $\pm$ 0.13 & ... & ... & ...
 & ... \\
 40 & 19:43:13.07 & 23:43:56.1 & ... & ... & 15.19 $\pm$ 0.16 & ... & ... & ...
 & ... \\
 41 & 19:43:10.81 & 23:43:56.2 & 16.91 $\pm$ 0.17 & 15.24 $\pm$ 0.16 & 
13.67 $\pm$ 0.11 & ... & ... & ... & ... \\
 42 & 19:43:10.12 & 23:43:58.0 & ... & ... & 14.28 $\pm$ 0.17 & ... & ... & ...
 & ... \\
 43 & 19:43:13.26 & 23:43:58.7 & ... & ... & 16.21 $\pm$ 0.17 & ... & ... & ...
 & ... \\
 44 & 19:43:11.44 & 23:43:59.1 & 17.30 $\pm$ 0.22 & 15.62 $\pm$ 0.16 & 
14.55 $\pm$ 0.12 & ... & ... & ... & ... \\
 45 & 19:43:09.38 & 23:44: 0.6 & 11.52 $\pm$ 0.08 & 11.10 $\pm$ 0.11 & 
10.87 $\pm$ 0.08 & 10.77 $\pm$ 0.07 & 10.76 $\pm$ 0.32 & ... & ... \\
 46 & 19:43:11.84 & 23:44: 1.4 & 16.44 $\pm$ 0.13 & 15.10 $\pm$ 0.15 & 
14.41 $\pm$ 0.12 & ... & ... & ... & ... \\
 47 & 19:43:13.09 & 23:44: 1.8 & ... & 15.63 $\pm$ 0.17 & 14.32 $\pm$ 0.14 & 
13.06 $\pm$ 0.17 & 12.11 $\pm$ 0.17 & ... & ... \\
 50 & 19:43:11.52 & 23:44: 4.6 & ... & ... & 14.66 $\pm$ 0.23 & ... & ... & ...
 & ... \\
 51 & 19:43:10.54 & 23:44: 4.3 & ... & ... & 15.43 $\pm$ 0.13 & ... & ... & ...
 & ... \\
 52 & 19:43:08.96 & 23:44: 4.7 & ... & 15.67 $\pm$ 0.16 & 14.19 $\pm$ 0.11 & 
13.48 $\pm$ 0.10 & 12.66 $\pm$ 0.33 & ... & ... \\
 53 & 19:43:08.28 & 23:44: 4.9 & 13.40 $\pm$ 0.14 & 12.38 $\pm$ 0.12 & 
12.18 $\pm$ 0.10 & 12.14 $\pm$ 0.10 & 12.14 $\pm$ 0.16 & ... & ... \\
 54 & 19:43:11.36 & 23:44: 5.3 & ... & ... & 13.46 $\pm$ 0.15 & ... & ... & ...
 & ... \\
 55 & 19:43:10.78 & 23:44: 5.3 & 14.91 $\pm$ 0.03 & 13.78 $\pm$ 0.16 & 12.21 $\pm$ 0.10 & ...
 & ... & ... & ... \\
 56 & 19:43:11.19 & 23:44: 4.3 & ... & ... & 11.21 $\pm$ 0.13 &  5.32 $\pm$ 0.17
 &  3.47 $\pm$ 0.22 &  3.53 $\pm$ 0.07 & ... \\
 57 & 19:43:08.41 & 23:44: 6.2 & ... & 17.02 $\pm$ 0.53 & 15.40 $\pm$ 0.17 & ...
 & ... & ... & ... \\
 58 & 19:43:13.43 & 23:44: 6.4 & ... & ... & 15.25 $\pm$ 0.13 & ... & ... & ...
 & ... \\
 59 & 19:43:11.46 & 23:44: 6.6 & ... & ... & 13.91 $\pm$ 0.17 & ... & ... & ...
 & ... \\
 60 & 19:43:10.14 & 23:44: 6.9 & ... & ... & 15.23 $\pm$ 0.13 & ... & ... & ...
 & ... \\
 61 & 19:43:11.74 & 23:44: 8.3 & ... & ... & 12.90 $\pm$ 0.16 & ... & ... & ...
 & ... \\
 62 & 19:43:12.00 & 23:44: 9.0 & ... & 15.37 $\pm$ 0.19 & 13.69 $\pm$ 0.15 & ...
 & ... & ... & ... \\
 63 & 19:43:10.77 & 23:44: 9.2 & 15.77 $\pm$ 0.13 & 15.36 $\pm$ 0.16 & 
14.95 $\pm$ 0.13 & ... & ... & ... & ... \\
 64 & 19:43:08.99 & 23:44: 9.4 & 14.09 $\pm$ 0.11 & 12.92 $\pm$ 0.12 & 
12.05 $\pm$ 0.10 & 10.21 $\pm$ 0.05 &  8.96 $\pm$ 0.05 &  7.74 $\pm$ 0.03 & 
 5.76 $\pm$ 0.03 \\
 65 & 19:43:08.47 & 23:44: 9.5 & ... & ... & 16.50 $\pm$ 0.20 & ... & ... & ...
 & ... \\
 66 & 19:43:13.03 & 23:44: 9.2 & 17.80 $\pm$ 0.26 & ... & 17.08 $\pm$ 0.30 & ...
 & ... & ... & ... \\
 67 & 19:43:11.48 & 23:44:10.6 & ... & ... & 15.22 $\pm$ 0.14 & ... & ... & ...
 & ... \\
 68 & 19:43:12.03 & 23:44:11.3 & ... & ... & 14.47 $\pm$ 0.14 & ... & ... & ...
 & ... \\
 69 & 19:43:10.06 & 23:44:11.4 & ... & ... & 15.02 $\pm$ 0.12 & ... & ... & ...
 & ... \\
 71 & 19:43:11.17 & 23:44:12.1 & ... & 14.37 $\pm$ 0.14 & 11.02 $\pm$ 0.09 & 
 8.10 $\pm$ 0.06 & ... &  5.37 $\pm$ 0.07 &  4.21 $\pm$ 0.11 \\
 72 & 19:43:13.27 & 23:44:12.3 & ... & 13.62 $\pm$ 0.13 & 11.27 $\pm$ 0.09 & 
 9.63 $\pm$ 0.08 &  9.32 $\pm$ 0.06 &  8.83 $\pm$ 0.28 & ... \\
 73 & 19:43:12.26 & 23:44:11.7 & ... & ... & 13.56 $\pm$ 0.16 & ... & ... & ...
 & ... \\
 74 & 19:43:11.60 & 23:44:13.4 & 17.63 $\pm$ 0.29 & 15.97 $\pm$ 0.17 & 
15.31 $\pm$ 0.13 & ... & ... & ... & ... \\
 75 & 19:43:08.90 & 23:44:13.6 & ... & ... & 16.48 $\pm$ 0.20 & ... & ... & ...
 & ... \\
 76 & 19:43:09.14 & 23:44:14.4 & 14.02 $\pm$ 0.10 & 13.41 $\pm$ 0.13 & 
12.85 $\pm$ 0.10 & 12.08 $\pm$ 0.07 & 11.60 $\pm$ 0.08 & 11.05 $\pm$ 0.12 & ...
 \\
 77 & 19:43:11.20 & 23:44:15.7 & ... & ... & 15.08 $\pm$ 0.15 & ... & ... & ...
 & ... \\
 78 & 19:43:10.80 & 23:44:15.4 & ... & ... & 16.85 $\pm$ 0.27 & ... & ... & ...
 & ... \\
 79 & 19:43:13.27 & 23:44:16.8 & ... & 16.57 $\pm$ 0.21 & 15.32 $\pm$ 0.16 & ...
 & ... & ... & ... \\
 80 & 19:43:11.09 & 23:44:16.8 & ... & ... & 15.14 $\pm$ 0.13 & ... & ... & ...
 & ... \\
 81 & 19:43:12.39 & 23:44:17.2 & ... & 16.65 $\pm$ 0.29 & 15.47 $\pm$ 0.15 & ...
 & ... & ... & ... \\
 82 & 19:43:12.57 & 23:44:17.9 & ... & 17.24 $\pm$ 0.29 & 16.58 $\pm$ 0.25 & ...
 & ... & ... & ... \\
 83 & 19:43:08.84 & 23:44:18.5 & ... & 15.84 $\pm$ 0.16 & 14.70 $\pm$ 0.12 & 
13.82 $\pm$ 0.12 & 13.53 $\pm$ 0.21 & ... & ... \\
 84 & 19:43:10.76 & 23:44:21.2 & 14.30 $\pm$ 0.10 & 14.02 $\pm$ 0.13 & 
13.75 $\pm$ 0.11 & 13.42 $\pm$ 0.18 & ... & 11.10 $\pm$ 0.24 & ... \\
 85 & 19:43:12.44 & 23:44:22.0 & ... & ... & 16.74 $\pm$ 0.23 & 13.55 $\pm$ 0.20
 & ... & ... & ... \\
 86 & 19:43:11.12 & 23:44:22.5 & ... & 16.39 $\pm$ 0.30 & 15.47 $\pm$ 0.13 & ...
 & ... & ... & ... \\
 87 & 19:43:13.10 & 23:44:23.6 & ... & ... & 16.48 $\pm$ 0.34 & ... & ... & ...
 & ... \\
 88 & 19:43:13.22 & 23:44:25.6 & 18.65 $\pm$ 0.32 & 16.77 $\pm$ 0.22 & 
16.08 $\pm$ 0.19 & ... & ... & ... & ... \\
 89 & 19:43:08.56 & 23:44:25.9 & ... & 15.16 $\pm$ 0.15 & 14.15 $\pm$ 0.11 & 
12.72 $\pm$ 0.09 & ... & ... & ... \\
 90 & 19:43:13.64 & 23:44:26.7 & ... & ... & 15.91 $\pm$ 0.25 & ... & ... & ...
 & ... \\
 91 & 19:43:13.45 & 23:44:29.5 & ... & ... & 15.40 $\pm$ 0.14 & 13.69 $\pm$ 0.14
 & 13.24 $\pm$ 0.13 & ... & ... \\
 92 & 19:43:11.07 & 23:44:29.9 & ... & ... & 16.12 $\pm$ 0.18 & ... & ... & ...
 & ... \\
 93 & 19:43:13.06 & 23:44:30.2 & ... & ... & 16.02 $\pm$ 0.19 & ... & ... & ...
 & ... \\
 94 & 19:43:09.25 & 23:44:32.9 & ... & ... & 13.78 $\pm$ 0.13 & ... & ... & ...
 & ... \\
 95 & 19:43:13.17 & 23:44:33.8 & ... & ... & 16.48 $\pm$ 0.21 & ... & ... & ...
 & ... \\
 96 & 19:43:10.81 & 23:44:34.0 & 15.94 $\pm$ 0.13 & 14.92 $\pm$ 0.15 & 
14.42 $\pm$ 0.12 & 14.03 $\pm$ 0.10 & ... & ... & ... \\
 97 & 19:43:09.42 & 23:44:34.4 & 13.36 $\pm$ 0.10 & 12.53 $\pm$ 0.12 & 
12.04 $\pm$ 0.10 & ... & ... & ... & ... \\
 98 & 19:43:08.72 & 23:44:36.0 & ... & ... & 15.74 $\pm$ 0.15 & ... & ... & ...
 & ... \\
 99 & 19:43:08.07 & 23:44:36.0 & 16.53 $\pm$ 0.15 & 15.35 $\pm$ 0.15 & 
14.89 $\pm$ 0.12 & 14.39 $\pm$ 0.12 & 14.32 $\pm$ 0.27 & ... & ... \\
100 & 19:43:08.94 & 23:44:36.4 & 14.90 $\pm$ 0.12 & 14.41 $\pm$ 0.15 & 
14.35 $\pm$ 0.12 & ... & ... & ... & ... \\
101 & 19:43:07.54 & 23:44:36.9 & ... & ... & 15.70 $\pm$ 0.13 & 15.11 $\pm$ 0.21
 & ... & ... & ... \\
102 & 19:43:12.78 & 23:44:36.7 & ... & ... & 17.56 $\pm$ 0.46 & ... & ... & ...
 & ... \\
103 & 19:43:08.89 & 23:44:37.4 & ... & ... & 14.32 $\pm$ 0.12 & ... & ... & ...
 & ... \\
104 & 19:43:11.87 & 23:44:37.6 & ... & ... & 15.50 $\pm$ 0.20 & 13.85 $\pm$ 0.11
 & 13.62 $\pm$ 0.18 & 12.05 $\pm$ 0.16 & ... \\
105 & 19:43:12.01 & 23:44:40.6 & ... & ... & 16.27 $\pm$ 0.20 & 14.07 $\pm$ 0.11
 & 13.05 $\pm$ 0.13 & ... & ... \\
106 & 19:43:12.93 & 23:44:41.0 & ... & ... & 15.21 $\pm$ 0.18 & 13.92 $\pm$ 0.07
 & 13.83 $\pm$ 0.17 & ... & ... \\
107 & 19:43:10.47 & 23:44:43.2 & ... & 16.86 $\pm$ 0.23 & 15.65 $\pm$ 0.15 & ...
 & ... & ... & ... \\
108 & 19:43:11.96 & 23:44:44.0 & ... & 16.26 $\pm$ 0.27 & 15.82 $\pm$ 0.15 & ...
 & ... & ... & ... \\
109 & 19:43:07.64 & 23:44:44.4 & 14.53 $\pm$ 0.11 & 13.81 $\pm$ 0.13 & 
13.57 $\pm$ 0.11 & 13.44 $\pm$ 0.08 & 13.33 $\pm$ 0.16 & ... & ... \\
110 & 19:43:10.04 & 23:44:45.2 & ... & 16.38 $\pm$ 0.18 & 15.97 $\pm$ 0.17 & ...
 & ... & ... & ... \\
111 & 19:43:09.35 & 23:44:46.2 & ... & 16.88 $\pm$ 0.21 & 15.54 $\pm$ 0.17 & ...
 & ... & ... & ... \\
112 & 19:43:12.61 & 23:44:46.9 & ... & ... & 17.07 $\pm$ 0.26 & 14.33 $\pm$ 0.12
 & 13.72 $\pm$ 0.17 & ... & ... \\
113 & 19:43:07.70 & 23:44:49.8 & 17.03 $\pm$ 0.18 & 16.04 $\pm$ 0.18 & 
15.63 $\pm$ 0.13 & 15.11 $\pm$ 0.17 & ... & ... & ... \\
114 & 19:43:11.49 & 23:44:50.4 & ... & ... & 15.64 $\pm$ 0.19 & 13.80 $\pm$ 0.12
 & 13.03 $\pm$ 0.14 & 11.97 $\pm$ 0.19 & 10.74 $\pm$ 0.13 \\
115 & 19:43:12.38 & 23:44:50.8 & ... & ... & 15.27 $\pm$ 0.19 & 13.34 $\pm$ 0.09
 & 13.98 $\pm$ 0.20 & ... & ... \\
116 & 19:43:12.50 & 23:44:52.2 & ... & 16.11 $\pm$ 0.22 & 14.96 $\pm$ 0.16 & ...
 & 13.65 $\pm$ 0.14 & ... & ... \\

\label{tbl:phot}
\end{longtable}

\end{landscape}

\end{document}